\documentclass[letterpaper,english,aps,prl,superscriptaddress,twocolumn, eqsecnum,amsfonts, amssymb]{revtex4}
\usepackage[T1]{fontenc}
\usepackage[latin9]{inputenc}
\usepackage{textcomp}
\usepackage{mathrsfs}
\usepackage{amsmath}
\usepackage{amssymb}
\usepackage{graphicx}
\usepackage{esint}

\makeatletter

\usepackage[bookmarks=true,colorlinks,linkcolor=blue,urlcolor=blue,citecolor=blue]{hyperref}

\newcommand{\be}{\begin{equation}}
\newcommand{\ee}{\end{equation}}
\newcommand{\bea}{\begin{eqnarray}}
\newcommand{\eea}{\end{eqnarray}}

\newcommand{\mc}{\mathcal}

\usepackage{epsfig,graphicx,psfrag,amsmath,amssymb,float}
\usepackage[caption=false]{subfig}

\@ifundefined{showcaptionsetup}{}{%
 \PassOptionsToPackage{caption=false}{subfig}}
\usepackage{subfig}
\makeatother

\usepackage{babel}

\begin{document}

\title{Dynamics of a $j=3/2$ quantum spin liquid}

\author{W. M. H. Natori}
\affiliation{Instituto de F\'{i}sica de S\~ao Carlos, Universidade de S\~ao Paulo, CP 369, S\~ao Carlos, SP, 13560-970, Brazil}
\author{M. Daghofer }
\affiliation{Institut f\"ur Funktionelle Materie und Quantentechnologien, Universit\"at Stuttgart, Keplerstra{\ss}e 7, 70174, Stuttgart, Germany}
\author{R. G. Pereira}
\affiliation{Instituto de F\'{i}sica de S\~ao Carlos, Universidade de S\~ao Paulo, CP 369, S\~ao Carlos, SP, 13560-970, Brazil}
\affiliation{
International Institute of Physics  and
Departamento de F\'isica Te\'orica e Experimental, Universidade Federal do Rio Grande do Norte, 59078-970 Natal-RN, Brazil}

\begin{abstract}
We study  a spin-orbital model for 4$d^{1}$ or 5$d^{1}$ Mott insulators
in ordered double perovskites with   strong 
spin-orbit coupling. This model is conveniently written
in terms of pseudospin and pseudo-orbital operators representing multipoles
of the effective $j=3/2$ angular momentum. Similarities between this model and  the  effective theories of Kitaev materials  motivate the proposal of a chiral spin-orbital  liquid with Majorana fermion excitations.
The thermodynamic and spectroscopic properties of this quantum spin liquid are characterized
using parton mean-field theory. The heat capacity, spin-lattice relaxation
rate, and dynamic structure factor for inelastic neutron scattering
are calculated and compared with the experimental data for the spin liquid   
candidate Ba$_{2}$YMoO$_{6}$. Moreover, based on a symmetry analysis, we discuss the operators involved in   resonant
inelastic X-ray scattering (RIXS) amplitudes  for  double perovskite
compounds. In general,  the RIXS cross sections  allow one to selectively   probe   pseudospin and pseudo-orbital degrees of freedom.  For the chiral spin-orbital liquid in particular, these cross sections provide information about the spectrum for different flavors of Majorana fermions. 
\end{abstract}
\maketitle

\section{Introduction\label{intro}}

Quantum spin liquids (QSLs) are highly entangled phases of matter
arising in strongly interacting spin systems \cite{Savary2016}.
Their intrinsic nonlocal character makes them elusive, since standard
experimental techniques probe two-point correlation functions. Experiments
performed on QSL candidates must then combine the outcomes of different
techniques with a careful theoretical analysis \cite{Savary2016,Lacroix2011,Avella2014}.
The difficulties to experimentally verify these quantum states of matter highlight
the importance of studying effective Hamiltonians which stabilize them 
as ground states. If one could calculate the response functions for  
these Hamiltonians, general properties of QSLs could be  investigated accurately, thus 
 guiding the design and interpretation of experiments. 

The Kitaev model on the honeycomb lattice plays an important role in this context \cite{Kitaev2006}.
This spin Hamiltonian displays Ising-like interactions along  different
quantization axes depending on the bond directions. This causes
an exchange frustration that drives the system to a Majorana QSL ground
state \cite{Baskaran2007}. Thanks to  its integrability, several
thermodynamic \cite{Nasu2014,Nasu2015,Nasu2016,Yoshitake2016} and
spectroscopic \cite{Knolle2014,Knolle2015,Smith2015,Smith2016,Halasz2016}
responses of the Kitaev model have been calculated exactly. 

Remarkably, the seminal work of Jackeli and Khaliullin  \cite{Jackeli2009} showed
that the Kitaev model is a good starting point to describe the magnetism
of certain $4d^{5}$ or 5$d^{5}$ Mott insulators.
To derive the Kitaev Hamiltonian, they considered the interplay of $t_{2g}$ orbital
directionality, hole virtual transfer through intermediate oxygen
orbitals, electronic correlation and strong spin-orbit coupling (SOC)
\cite{Khaliullin2005}. The work in Ref. \cite{Jackeli2009} made specific proposals
for candidate materials  that could exhibit a QSL ground state, leading to a
manifold of experimental studies, exemplified by Refs. \cite{SinghPRL2012,PlumbPRB2014,TakayamaPRL2015,BanerjeeNatMat2016}.
Unfortunately, none of the  compounds studied so far harbors   a Majorana QSL, showing  instead different types of magnetic order at low temperatures.   
The magnetic order in these materials can be explained by the effects of competing exchange interactions which have to be added to the Kitaev model \cite{Chaloupka2010,Rau2014,Winter2016}.
The 
effective Hamiltonians generated by the Jackeli-Khaliullin mechanism are examples of 
quantum compass models, which are known to host 
unusual magnetism \cite{Nussinov2015}. The wealth  of theoretical proposals and experiments has led 
much  of the research on QSLs  to turn to  compounds
that combine strong correlations and SOC \cite{Witczak-Krempa2014,Schaffer2015,Winter2017}.

\begin{figure}
\begin{centering}
\subfloat[\label{fig:ODP}]{\begin{centering}
\includegraphics[width=0.45\columnwidth]{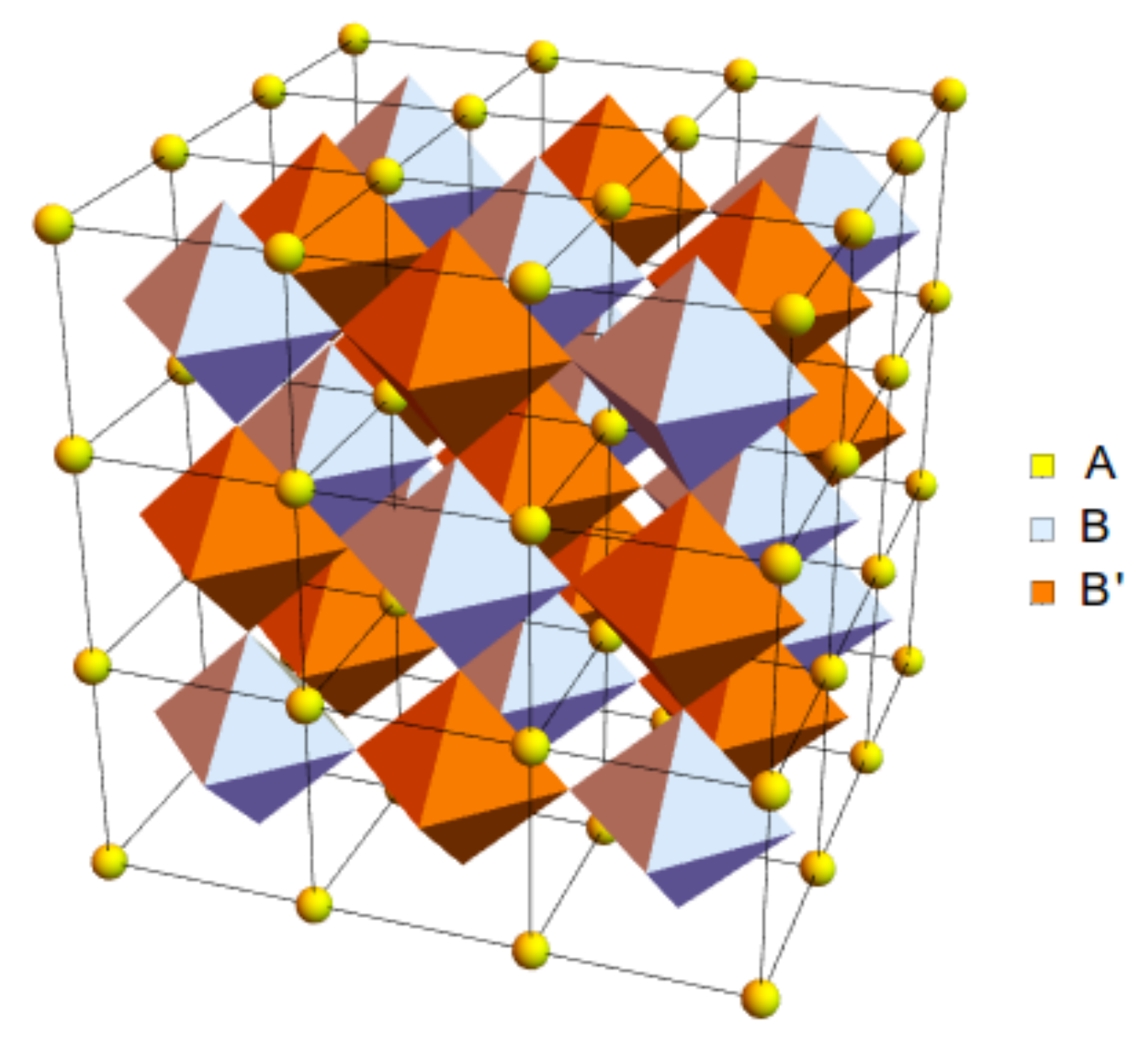}
\par\end{centering}

}\subfloat[\label{fig:bond-dependent exchange}]{\includegraphics[width=0.45\columnwidth]{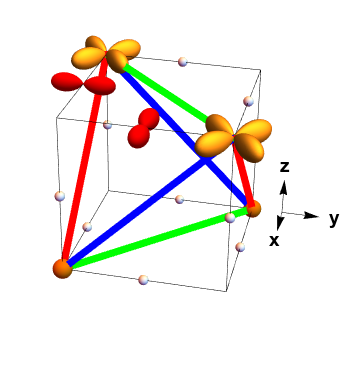}

}
\par\end{centering}

\begin{raggedright}
\caption{(Color online) (\ref{fig:ODP}) Crystal structure of ordered double
perovskites, with chemical formula A$_{2}$BB'O$_{6}$. The oxygen
$O^{2-}$ anions correspond to the vertices of the octahedra. (\ref{fig:bond-dependent exchange})
Tetrahedron of the B' species, highlighting the exchange path. Different bond colors represent different first-neighbor
interactions in $xy$, $yz$ or $xz$ planes.}

\par\end{raggedright}

\end{figure}

Mott insulators in ordered double perovskites based on heavy $d$
ions satisfy the conditions leading to quantum compass models. Ordered
double perovskites are oxides of general stoichiometry A$_{2}$BB'O$_{6}$,
where A corresponds to an alkaline-earth or lanthanide, and B, B' are
transition metal ions (Fig. \ref{fig:ODP}). Chen \emph{et al.}
\cite{Chen2010} put forward  a spin-orbital model for compounds in
which B' is the only magnetically  active ion in a 4$d^{1}$ or 5$d^{1}$
electronic configuration. In  materials that retain cubic symmetry, the  spin and orbital angular momenta of the electron in the $t_{2g}$ orbital combine to form  an effective $j=3/2$ magnetic moment. 
The effective spin Hamiltonian in this case contains bond-dependent anisotropic interactions between $j=3/2$ moments distributed on an fcc lattice (Fig. \ref{fig:bond-dependent exchange}).
However, in contrast to the  Kitaev model, the interactions involve higher 
  multipoles of the   angular momentum. 

The combination of geometric frustration in the fcc lattice and multipolar interactions induced by SOC can favor exotic phases such as  valence bond solids or QSLs \cite{Chen2010}.
In coherence with these predictions, experimental results show that
the double perovskite Ba$_{2}$YMoO$_{6}$ does not present any structural
transition or magnetic order down to 2K (much lower than its Curie-Weiss
temperature) \cite{deVries2010,Aharen2010,Carlo2011,deVries2013}.
Motivated by these observations,  a chiral spin-orbital liquid has been proposed as a possible ground state  of the double perovskite  model in a particular regime \cite{Natori2016}. This
QSL is similar to some three-dimensional versions of the Kitaev model
\cite{Hermanns2015,OBrien2016}, as it exhibits Majorana fermion
excitations with a gapless nodal-line spectrum  instead of a Fermi surface. Another theoretical proposal to explain the properties  of Ba$_{2}$YMoO$_{6}$  is the disordered dimer-singlet phase  \cite{Romhanyi2017}. The latter shares  with the chiral spin-orbital liquid the property  of pseudo-gapped low-energy excitations, which are however   due to a random distribution of dimerized bonds. 

In this work, we calculate  various response functions  of the $j=3/2$ chiral spin-orbital
liquid proposed in Ref. \cite{Natori2016} within the mean-field approximation of free Majorana fermions. We calculate the magnetic specific heat, spin-lattice relaxation
rate, and inelastic neutron scattering (INS) cross section  and compare the theoretical
results with the experimental data   for Ba$_{2}$YMoO$_{6}$ \cite{deVries2010,Aharen2010,Carlo2011,deVries2013}.
We also investigate the expected resonant inelastic X-ray scattering
(RIXS)  cross sections of the chiral spin-orbital liquid. RIXS has grown in importance in recent years  \cite{Haverkort2010,Ament2011},   due to its ability
of probing orders that are hidden from neutron experiments \cite{Savary2015}
and of distinguishing different types of excitations by tuning the
polarization and energy of the photons. In fact, recent studies of
RIXS cross sections for the Kitaev honeycomb model 
showed that they can probe gapless Majorana fermions and gapped visons
separately \cite{Halasz2016}. This result is in sharp contrast to  the dynamic structure factor measured in neutron scattering,  which probes 
  spin excitations that in the Kitaev model always excite  a gapped vison \cite{Knolle2014,Knolle2015,Smith2015,Smith2016}.
Therefore, RIXS can give more information about Majorana fermions
than the usual experiments. This is particularly interesting for the chiral spin-orbital liquid \cite{Natori2016}, whose spectrum contains  non-degenerate flavors of Majorana fermions associated with different spin-orbital excitations.

The paper is organized as follows. In Section \ref{sec:Model-and-Symmetry}, we describe 
the electronic structure of the Mo$^{5+}$ ion, relevant for the magnetism in  Ba$_{2}$YMoO$_{6}$,   and derive the 
microscopic Hamiltonian for ordered double perovskites.
Section \ref{sec:QSL-MFT} discusses the parton mean-field
theory for  the chiral spin-orbital liquid state. In Section
\ref{sec:usual probes}, we present our results for specific heat,
spin-lattice relaxation rate and INS cross section, providing comparison
with available experimental data.  Section \ref{sec:RIXS} contains our results for   RIXS scattering operators for ordered double perovskites, based on a symmetry analysis of the $L$ absorption edge. These results apply in general to   4$d^{1}$ and 5$d^{1}$ based compounds. An important
outcome of this analysis is a proposal of how to directly probe pseudospin and pseudo-orbital  degrees of freedom of the $j=3/2$ multiplet. 
We apply these results  in particular to calculate  the RIXS   cross sections of the chiral spin-orbital liquid.  Finally,
we summarize our results and suggest future developments for theory
and experiments in Sec. \ref{sec:Conclusions}. Technical details
of the calculations and complementary   results   are left to the Appendices. 

\section{Model and Symmetry \label{sec:Model-and-Symmetry}}

\subsection{$t_{2g}$ orbitals in $d^{1}$ configuration\label{sub:ionic-gs} }

We start by  discussing  the orbital physics of singly occupied $t_{2g}$ orbitals.
Double perovskites with stoichiometry A$_{2}$BB'O$_{6}$ are structurally
formed by corner-sharing BO$_{6}$ and B'O$_{6}$ octahedra, arranged
as shown in Fig. \ref{fig:ODP}. The projection of the angular momentum
$\mathbf{L}$ ($L=2$ for $d$ orbitals) onto the $t_{2g}$ triplet defines a $l=1$
effective angular momentum $\mathbf{l}$ \cite{Fazekas1999,khomskii2014transition}:
\begin{equation}
\textbf{l}=-\mathcal{P}_{t_{2g}}\textbf{L}\mathcal{P}_{t_{2g}},
\end{equation}
in which $\mathcal{P}_{t_{2g}}$ is the projection operator. Let $d_{\alpha\beta,\sigma}$
be the annihilation operator for an electron in the $\alpha\beta$ orbital
$(\alpha\beta\in\{xy,yz,xz\})$ with spin $\sigma\in\{\uparrow,\downarrow\}$,
and $d_{m_{l},\sigma}$ the corresponding operators for eigenstates
of $l^{z}$, with eigenvalue $m_{l}\in\{-1,0,1\}$. The relation 
between these operators is     \cite{Chen2010}
\begin{align}
d_{0,\sigma}&=d_{xy,\sigma},\label{dfromalphatom1} \\
 d_{\pm1,\sigma}&=\frac{\mp d_{yz,\sigma}-id_{zx,\sigma}}{\sqrt{2}}.\label{dfromalphatom2}
\end{align}

Equations  (\ref{dfromalphatom1}) and (\ref{dfromalphatom2}) provide a complete basis to describe the
physics of $d^{1}$ strongly correlated systems. Spin-orbital models 
for double perovskites considering all states spanned by this basis
were studied in Ref.  \cite{Svoboda2017}. These general models interpolate 
between the weak and strong SOC limits. 
Here we focus on the limit  in which the SOC is strong enough to
justify a projection of the Hamiltonian onto a low-energy subspace.  The ionic spin-orbit Hamiltonian is written as
\begin{equation}
H_{\text{ion}}=-\lambda\textbf{\textbf{l}}\cdot\textbf{S},\label{eq:Hion}
\end{equation}
in which $\mathbf{S}$ is the electronic spin and $\lambda>0$ is the SOC constant. The effect of $H_{\text{ion}}$
is to split  the $t_{2g}$ levels into one $j=1/2$ and one $j=3/2$ manifold
($\textbf{J}=\textbf{l}+\textbf{S}$), the latter being energetically
favored by a gap of $3\lambda/2$.

It is convenient to organize the six eigenstates of $H_{\text{ion}}$
into three Kramers' doublets. 
We define the corresponding annihilation operators  by \cite{Kim2012,Chaloupka2016}
\begin{subequations}\label{eq:ABC}
\begin{align}
A_{\sigma} & =2\sigma\left(\frac{1}{\sqrt{3}}\,d_{0,-\sigma}-\sqrt{\frac{2}{3}}\,d_{-2\sigma,\sigma}\right),  \\
B_{\sigma} & =\sqrt{\frac{2}{3}}\,d_{0,-\sigma}+\frac{1}{\sqrt{3}}\,d_{-2\sigma,\sigma},  \\
C_{\sigma} & =d_{2\sigma,\sigma}.
\end{align}
\end{subequations}
where  $\sigma=\uparrow,\downarrow=\pm 1/2$   distinguishes between Kramers-degenerate states. Note that in this notation the index $\sigma$ in $A_\sigma$ and $B_\sigma$ is not directly connected with the actual spin eigenvalue in the $d$ operators on the right-hand side of Eqs. (\ref{eq:ABC}). 

\begin{figure}
\begin{centering}
\includegraphics[width=0.85\columnwidth]{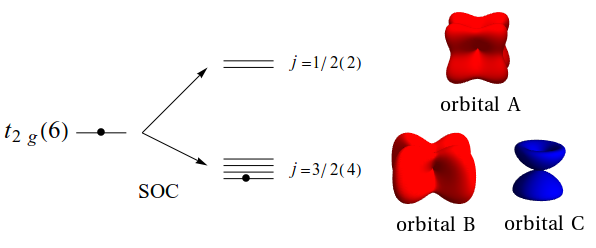}
\par\end{centering}

\caption{\label{fig:orbital sketches}Energy level splitting
of 4$d^{1}$ or 5$d^{1}$ electrons in the presence of a  cubic-symmetric crystal field and   spin-orbit coupling. The   density profiles of the $A$, $B$ and $C$ states are also illustrated. }
\end{figure}

Figure \ref{fig:orbital sketches} shows  the level splitting and the electronic density profiles of the
$A$, $B$ and $C$ states. The $A$ states are associated with the higher-energy $j=1/2$ subspace. In the presence of cubic symmetry (which is the case for  Ba$_{2}$YMoO$_{6}$  \cite{Chen2010}), the $B$ and $C$ states are degenerate and form the $j=3/2$ multiplet. However, since they have different electronic distributions, their degeneracy would be lifted by a tetragonal lattice distortion (see Appendix \ref{sec:Orbital-Physics}). 
 In terms of eigenstates $|j,m_j\rangle$ of $\mathbf J^2$ and $J^z$, we can identify the states created by $B_{\sigma}^\dagger$ and $C_{\sigma}^\dagger$ as
 \begin{subequations}\label{eq:relation states}
\begin{align}
\left|B_{\uparrow}\right\rangle  & =\left|\frac{3}{2},\frac{1}{2}\right\rangle,  \\
\left|B_{\downarrow}\right\rangle  & =\left|\frac{3}{2},-\frac{1}{2}\right\rangle  ,  \\
\left|C_{\uparrow}\right\rangle  & =\left|\frac{3}{2},\frac{3}{2}\right\rangle ,  \\
\left|C_{\downarrow}\right\rangle  & =\left|\frac{3}{2},-\frac{3}{2}\right\rangle  .
\end{align}
\end{subequations}

Alternatively, we can use two pseudospins $1/2$ to label  the four states in the $j=3/2$ subspace \cite{Natori2016}. The first pseudospin is denoted $s$ and is associated with the Kramers degeneracy:\begin{align}
s= & \begin{cases}
-\sigma, & \text{for }B_\sigma,\\
+\sigma, & \text{for }C_\sigma.
\end{cases}
\end{align}
The second pseudospin, hereafter called pseudo-orbital   $\tau$, is defined by  
\begin{align}
\tau= & \begin{cases}
-\frac{1}{2}, & \text{for }B,\\
+\frac{1}{2}, & \text{for }C.
\end{cases}\label{eq:pseudoorbital definition}
\end{align}
In the notation of $|s,\tau\rangle$, with $s,\tau=\pm 1/2$, we   write
\begin{subequations}\label{eq:relation states2}
\begin{align}
\left|B_{\uparrow}\right\rangle  &  =\left|-\frac12,-\frac12\right\rangle ,  \\
\left|B_{\downarrow}\right\rangle  &  =\left|\frac12,-\frac12\right\rangle ,  \\
\left|C_{\uparrow}\right\rangle  &  =\left|\frac12,\frac12\right\rangle ,  \\
\left|C_{\downarrow}\right\rangle  &   =\left|-\frac12,\frac12\right\rangle .
\end{align}
\end{subequations}
This definition is such that the $z$ component of the total angular momentum    is given by
\begin{equation}
J^{z}=s^{z}+4s^{z}\tau^{z},\label{eq:Jz}
\end{equation}
where the operators $s^z$ and $\tau^z$ are defined by \bea
s^z|s,\tau\rangle&=&s|s,\tau\rangle, \\
\tau^z|s,\tau\rangle&=&\tau|s,\tau\rangle. \eea
More generally,
if we define the vector of annihilation operators $\xi\equiv(
C_{\uparrow} , C_{\downarrow} , B_{\uparrow} , B_{\downarrow})^{t}$, we have a basis of operators in the space of a singly occupied $j=3/2$ level: 
\begin{align}
s^{a} & =\frac{1}{2}\xi^{\dagger}(\mathbb{I}\otimes\sigma^{a})\xi,  \\
\tau^{a} & =\frac{1}{2}\xi^{\dagger}(\sigma^{a}\otimes\mathbb{I})\xi,  \\
s^{a}\tau^{b} & =\frac{1}{4}\xi^{\dagger}(\sigma^{b}\otimes\sigma^{a})\xi,
\end{align}
where $\sigma^a$, with $a\in\{x,y,z\}$, are Pauli matrices and $\mathbb I$ is the $2\times2$ identity matrix. 
The pseudospin and pseudo-orbital  operators obey the SU(2) algebra:
\begin{align}
[s^{a},s^{b}]&=i\epsilon^{abc}s^c,\label{Su2s}\\
[\tau^{a},\tau^{b}]&=i\epsilon^{abc}\tau^c,\\
[s^{a},\tau^{b}]&=0\label{comust}.
\end{align}
Similar spin and orbital operators appear in Kugel-Khomskii models  for $e_{g}$ orbitals without SOC \cite{Feiner1997,PhysRevB.56.R14243,Li1998,Joshi1999,Wang2009,Corboz2012,Lajko2013}. A crucial    difference is that   here $\mathbf s$ and $\boldsymbol \tau$ act on $j=3/2$ states, in which spin and orbital degrees of freedom are entangled as described  by Eqs. (\ref{eq:ABC}).

\begin{table*}
\caption{\label{tab:multipoles}Operators   describing active multipoles
within a cubic $\Gamma_{8}$ quartet. Bars over functions of $\mathbf J$ indicate the
symmetrization with respect to all the possible permutations of the
indices, e.g., $\overline{J_{x}J_{y}^{2}}=J_{x}J_{y}^{2}+J_{y}J_{x}J_{y}+J_{y}^{2}J_{x}$.
Adapted from Refs. \cite{Santini2009,Shiina1998}.}
\begin{tabular}{cccc}
\hline 
Moment  & Symmetry  & $\textbf{J}$ multipoles  & $(\textbf{s},\boldsymbol{\tau})$ multipoles \tabularnewline
\hline 
Dipoles  & $\Gamma_{4}$  & $J^{x}$  & $s^{x}(1+4\tau^{yz})$ \tabularnewline
 &  & $J^{y}$  & $-s^{y}(1+4\tau^{xz})$ \tabularnewline
 &  & $J^{z}$  & $s^{z}(1+4\tau^{xy})$ \tabularnewline
\hline 
Quadrupoles  & $\Gamma_{3}$  & $\begin{aligned}O_{3z^{2}-r^{2}} & =3(J^{z})^{2}-\textbf{J}^{2}\equiv\hat{O}_{2}^{0}\end{aligned}
$  & $6\tau^{z}$ \tabularnewline
 &  & $O_{x^{2}-y^{2}}=(J^{x})^{2}-(J^{y})^{2}\equiv\hat{O}_{2}^{2}$  & $2\sqrt{3}\tau^{x}$ \tabularnewline
\hline 
 & $\Gamma_{5}$  & $O_{xy}=\frac{1}{2}\overline{J^{x}J^{y}}\equiv\hat{O}_{2}^{-2}$  & $2\sqrt{3}s^{z}\tau^{y}$ \tabularnewline
 &  & $O_{yz}=\frac{1}{2}\overline{J^{y}J^{z}}\equiv\hat{O}_{2}^{-1}$  & $2\sqrt{3}s^{x}\tau^{y}$ \tabularnewline
 &  & $O_{xz}=\frac{1}{2}\overline{J^{x}J^{z}}\equiv\hat{O}_{2}^{1}$  & $-2\sqrt{3}s^{y}\tau^{y}$ \tabularnewline
\hline 
Octupoles  & $\Gamma_{2}$  & $T_{xyz}=\frac{\sqrt{15}}{6}\overline{J^{x}J^{y}J^{z}}$  & $\frac{3\sqrt{5}}{2}\tau^{y}$ \tabularnewline
\hline 
 & $\Gamma_{4}$  & $T_{x}^{\alpha}=(J^{x})^{3}-\frac{1}{2}(\overline{J^{x}(J^{y})^{2}}+\overline{(J^{z})^{2}J^{x}})$  & $3s^{x}(1-\tau^{yz})$ \tabularnewline
 &  & $T_{y}^{\alpha}=(J^{y})^{3}-\frac{1}{2}(\overline{J^{y}(J^{z})^{2}}+\overline{(J^{x})^{2}J^{y}})$  & $-3s^{y}(1-\tau^{xz})$ \tabularnewline
 &  & $T_{z}^{\alpha}=(J^{z})^{3}-\frac{1}{2}(\overline{J^{z}(J^{x})^{2}}+\overline{(J^{y})^{2}J^{z}})$  & $3s^{z}(1-\tau^{xy})$ \tabularnewline
\hline 
 & $\Gamma_{5}$  & $T_{x}^{\beta}=\frac{\sqrt{15}}{6}[\overline{J^{x}(J^{y})^{2}}-\overline{(J^{z})^{2}J^{x}}]$  & $3\sqrt{5}s^{x}\overline{\tau}^{yz}$ \tabularnewline
 &  & $T_{y}^{\beta}=\frac{\sqrt{15}}{6}[\overline{J^{y}(J^{z})^{2}}-\overline{(J^{x})^{2}J^{y}}]$  & $-3\sqrt{5}s^{y}\overline{\tau}^{xz}$ \tabularnewline
 &  & $T_{z}^{\beta}=\frac{\sqrt{15}}{6}[\overline{J^{z}(J^{x})^{2}}-\overline{(J^{y})^{2}J^{z}}]$  & $3\sqrt{5}s^{z}\overline{\tau}^{xy}$ \tabularnewline
\hline 
\end{tabular}
\end{table*}

Table \ref{tab:multipoles} shows how the multipoles of $\textbf{J}$
are  written in terms of the components of $\textbf{s}$ and $\boldsymbol{\tau}$. Here we  introduce the linear combinations of $\tau^x$ and $\tau^z$:\begin{subequations}\label{taualphabeta}
\begin{align}
\tau^{xy}&=\tau^{z},\\
 \tau^{yz(zx)}&=\frac{1}{2}(-\tau^{z}\pm\sqrt{3}\tau^{x}),\\
\overline{\tau}^{xy}&=\tau^{x},\\
\overline{\tau}^{yz(zx)}&=-\frac{1}{2}(\tau^{x}\pm\sqrt{3}\tau^{z}).
\end{align}
\end{subequations}
According to Table \ref{tab:multipoles}, $\textbf{s}$ is a linear
combination of dipole and octupole moments of $\mathbf J$ in the $\Gamma_{4}$
representation. Similarly, $\tau^{x}$ and $\tau^{z}$ correspond 
  to quadrupoles in the $\Gamma_{3}$ representation. The
component $\tau^{y}$ appears separately as a one-dimensional representation $\Gamma_{2}$. We conclude that all the components of $\textbf{s}$  are odd under conjugation by  the
time-reversal operator $T$. As for  the pseudo-orbital $\boldsymbol{\tau}$, the $\tau^x$ and $\tau^z$ components are even while 
  $\tau^{y}$   is odd under time reversal. More explicitly,
\begin{equation}
T^{-1}\boldsymbol{\tau}T=(\tau^{x},-\tau^{y},\tau^{z}).
\end{equation}

\subsection{Interacting spin model\label{sub:Microscopic-Model}}

In this subsection, we reproduce the derivation of  the effective spin model for $d^1$ double perovskites following Ref.  \cite{Romhanyi2017}. 
We present this derivation here  for completeness and to mention some important
aspects in the interpretation of the model parameters.

We start from the multi-orbital Hubbard model \begin{align}
H =&-t\underset{\langle i,j\rangle_{\gamma},\sigma}{\sum}(d_{i,\alpha\beta,\sigma}^{\dagger}d_{j,\alpha\beta,\sigma}+\text{h.c.})\nonumber\\
&+U\underset{i,a}{\sum}n_{i,\alpha\beta,\uparrow}n_{i,\alpha\beta,\downarrow}-J_{H}\underset{i,\sigma}{\sum}\underset{\gamma\delta<\alpha\beta}{\sum}n_{i,\alpha\beta,\sigma}n_{i,\gamma\delta,\sigma}\nonumber \\
 & +(U-2J_{H})\underset{i,\sigma,\sigma^{\prime}}{\sum}\underset{\gamma\delta<\alpha\beta}{\sum}n_{i,\alpha\beta,\sigma}n_{i,\gamma\delta,\sigma^{\prime}}\nonumber \\
 & -J_{H}\underset{i,\gamma\delta<\alpha\beta}{\sum}\left(d_{i,\alpha\beta,\uparrow}^{\dagger}d_{i,\alpha\beta,\downarrow}d_{i,\gamma\delta,\downarrow}^{\dagger}d_{i,\gamma\delta,\uparrow}+\text{h.c.}\right)\nonumber\\
&-J_H\sum_{i,\gamma\delta<\alpha\beta} \left(d_{i,\alpha\beta,\uparrow}^{\dagger}d_{i,\gamma\delta,\downarrow}d_{i,\alpha\beta,\downarrow}^{\dagger}d_{i,\gamma\delta,\uparrow}+\text{h.c.}\right).
\end{align}
Here $i$ labels the lattice sites, $\alpha\beta$ labels the $t_{2g}$
orbitals,  $\sigma$ is the electronic spin, $U$ is the Coulomb interaction, and
$J_{H}$ is   Hund's coupling. We use the ordering convention $xy<yz<zx$.  The hopping processes are  restricted to nearest-neighbor sites, as
in Ref. \cite{Chen2010}, such that  $\gamma$ labels the axis perpendicular to the $\alpha\beta$  plane of the
$\langle i,j\rangle$  bond  (see
Fig. \ref{fig:bond-dependent exchange}).

Using the single-occupancy constraint  $\underset{\alpha\beta}{\sum}n_{i,\alpha\beta}=1$ and applying perturbation theory in the regime  $t\ll U,J_H$, we obtain
the  spin-orbital model \cite{Romhanyi2017}
\begin{align}
H_{\text{so}}  =&J\underset{\langle ij\rangle_{\gamma}}{\sum}\left(\textbf{S}_{i}\cdot\textbf{S}_{j}+\frac{1}{4}\right)n_{i,\alpha\beta}n_{j,\alpha\beta}\nonumber \\
 & -J^{\prime}\underset{\langle ij\rangle_{\gamma}}{\sum}\textbf{S}_{i}\cdot\textbf{S}_{j}P_{ij}^{(\gamma)} +\frac{3}{2}J^{\prime}\underset{\langle ij\rangle_{\gamma}}{\sum}n_{i,\alpha\beta}n_{j,\alpha\beta}\nonumber\\
&+V\underset{\langle ij\rangle_{\gamma}}{\sum}n_{i,\alpha\beta}n_{j,\alpha\beta},\label{eq:Heff without SOC}
\end{align}
in which $P_{ij}^{(\gamma)}=n_{i,\alpha\beta}\bar{n}_{j,\alpha\beta}+\bar{n}_{i,\alpha\beta}n_{j,\alpha\beta}$
with $\bar{n}_{i,\alpha\beta}=n_{i,\beta\gamma}+n_{i,\gamma\alpha}=1-n_{i,\alpha\beta}$.
The coupling constants $J$, $J^{\prime}$ and $V$   are   given by
\begin{align}
J & =\frac{K}{3}\left(2r_{3}+r_{2}\right),\label{eq:Jcoupling}\\
J^{\prime} & =\frac{K}{4}\left(r_{1}-r_{2}\right),\label{eq:Jpcoupling}\\
V & =\frac{K}{3}\left(r_{2}-r_{3}\right),\label{eq:Vcoupling}
\end{align}
where $K=4t^{2}/U$, $r_{1}=1/(1-3\eta)$, $r_{2}=1/(1-\eta)$ and
$r_{3}=1/(1+2\eta)$, with $\eta =J_H/U$. 

Equation  (\ref{eq:Heff without SOC}) can be  compared   with the
model derived in Ref. \cite{Chen2010}. The first line corresponds to
the antiferromagnetic exchange interaction, with a correction in the 
sign of the spin-independent term. The second line
is formally the ferromagnetic Hamiltonian; the difference lies in
the interpretation of the parameter $J^{\prime}$. We find  that
$J'$ is related with the ratio $J_{H}/U$ of the transition metal, instead of 
  the ratio at the oxygen site. The third line is   similar
to the electric quadrupole interaction discussed in Ref. \cite{Chen2010},
differing by the absence of a term proportional to $(n_{i,\beta\gamma}-n_{i,\gamma\alpha})(n_{j,\beta\gamma}-n_{j,\gamma\alpha})$.
In summary, our minimal model also contains antiferromagnetic, quadrupole
and ferromagnetic interactions. However, the explanation of the model
parameters comes from a different mechanism. 

Hereafter we focus on the limit of vanishing Hund's coupling $\eta=0$, in which $J=K$ and $J'=V=0$. This corresponds to the regime in which  quantum fluctuations are maximized and may favor a QSL ground state \cite{Chen2010}. The final step is the projection
of $H_{\text{so}}$ onto the $j=3/2$ manifold in the limit $\lambda\gg K$:\begin{equation}
 H_{\text{eff}}=\mathcal{P}_{3/2}H_{\text{so}}\mathcal{P}_{3/2}.\label{eq:projection}
\end{equation}
Using the pseudospins
and pseudo-orbitals discussed in Subsection \ref{sub:ionic-gs}, we
find for  $\eta=0$ \cite{Natori2016,Romhanyi2017}:
\begin{equation}
H_{\text{eff}}=\frac{4J}{9}\underset{\langle ij\rangle_{\gamma}}{\sum}\left(\textbf{s}_{i}\cdot\textbf{s}_{j}+\frac{1}{4}\right)\left(\frac{1}{2}-\tau_{i}^{\alpha\beta}\right)\left(\frac{1}{2}-\tau_{j}^{\alpha\beta}\right).\label{eq:main Hamiltonian}
\end{equation}
The bond-dependent exchange processes are represented in Fig. \ref{fig:bond-dependent exchange}.
In analogy with the effective models for iridates \cite{Jackeli2009,Chaloupka2010,Rau2014},
the anisotropy arises from the directionality of the $t_{2g}$ orbitals
\cite{Khaliullin2005}. In this notation, the  hidden SU(2)  symmetry of the effective model 
discussed in Ref. \cite{Chen2010} becomes transparent. More explicitly,
if we define the total pseudospin operator
\begin{equation}
\textbf{s}_{\text{tot}}=\underset{i}{\sum}\textbf{s}_{i},\label{eq:total pseudospin}
\end{equation}
then $[ H_{\text{eff}},\textbf{s}_{\text{tot}}]=0$. This continuous symmetry, unexpected for general spin-orbit coupled systems, enhances quantum fluctuations and favors unconventional magnetic states \cite{Chen2010}. In addition, a $Z_{3}$ symmetry corresponding to a $2\pi/3$ rotation of the $(\tau^z,\tau^x)$ vector  and analogous to the symmetry of quantum compass models \cite{Nussinov2015} is   made
evident by the $\tau^{\alpha\beta}$ pseudo-orbital operators.    These
symmetry properties  of $H_{\text{eff}}$   play an important role in the
ansatz for  the chiral spin-orbital liquid to be discussed  in Sec. \ref{sec:QSL-MFT}. 
The expression for the more general projected Hamiltonian with $\eta\neq0$ is  given in Appendix \ref{sec:Hund-Coupling-Terms}.

\section{Mean-Field Theory of the Chiral Spin-Orbital Liquid\label{sec:QSL-MFT}}

In the following we describe the parton mean-field theory that gives
rise to the chiral spin-orbital liquid studied in Ref. \cite{Natori2016}.
The motivation for considering  a Majorana fermion parton construction  arises mainly from the
similarities between the double perovskite model and the Kitaev model, as they both contain bond-dependent anisotropic exchange interactions. The main point of this section is to discuss the spectrum of Majorana fermion excitations, which will be important to interpret the response functions discussed in  Secs. \ref{sec:usual probes} and \ref{sec:RIXS}. 

The operators $\textbf{s}$ and $\boldsymbol{\tau}$ obeying the algebra in Eqs. (\ref{Su2s}) through (\ref{comust}) can be represented
by Majorana fermions in the following way \cite{PhysRevLett.69.2142,Coleman1994,Biswas2011,Herfurth2013,Wang2009}
\begin{align}
s^{a} & =-\frac{i}{4}\epsilon^{abc}\eta^{b}\eta^{c},\nonumber \\
\tau^{a} & =-\frac{i}{4}\epsilon^{abc}\theta^{b}\theta^{c},\label{eq:Majorana rep}
\end{align}
where  $a=x,y,z=1,2,3$ for the Majorana fermion flavors. 
The six Majorana fermions $\zeta^{a}\in\{\eta^{a},\theta^{a}\}$
obey $(\zeta^a)^\dagger=\zeta^a$ and $\left\{ \zeta^{a},\zeta^{b}\right\} =2\delta^{ab}$.
This representation has a $Z_{2}$ gauge structure because the sign
of the fermions can be changed ($ \eta^a\rightarrow- \eta^a$
and $ \theta^a\rightarrow- \theta^a$) without
modifying the local physical operators. Since the   Hilbert space is enlarged,
one needs to impose a local constraint at each site $j$ to identify 
the physical states: 
\begin{equation}
i\eta_{j}^{1}\eta_{j}^{2}\eta_{j}^{3}\theta_{j}^{1}\theta_{j}^{2}\theta_{j}^{3}=1.\label{eq:Majorana constraint}
\end{equation}
The local constraint also implies that \be
s^{a}\tau^{b}=-\frac{i}4\eta^{a}\theta^{b}.\ee

Using Eq. (\ref{eq:Majorana rep}), we rewrite the Hamiltonian in Eq. (\ref{eq:main Hamiltonian})
as 
\begin{align}
H_{\text{eff}} =&\frac{J}{36}\underset{\langle i,j\rangle_{\gamma}}{\sum}\left[\underset{a<b}{\sum}\eta_{i}^{a}\eta_{j}^{a}\eta_{i}^{b}\eta_{j}^{b}\right.\nonumber \\
 & +(\eta_{i}^{2}\eta_{i}^{3}\eta_{j}^{1}+\eta_{i}^{3}\eta_{i}^{1}\eta_{j}^{2}+\eta_{i}^{1}\eta_{i}^{2}\eta_{j}^{3})\bar{\theta}_{j}^{\alpha\beta}+(i\leftrightarrow j)\nonumber \\
 & \left.+\boldsymbol{\eta}_{i}\cdot\boldsymbol{\eta}_{j}\bar{\theta}_{i}^{\alpha\beta}\bar{\theta}_{j}^{\alpha\beta}+\theta_{i}^{\alpha\beta}\theta_{j}^{\alpha\beta}\theta_{i}^{2}\theta_{j}^{2}\right]+\text{const.}.\label{eq:Majorana main Hamiltonian}
\end{align}
Here we have introduced the fermions $\theta^{\alpha\beta}$ and $\bar\theta^{\alpha\beta}$ as linear combinations of $\theta^1$ and $\theta^3$ in analogy with Eqs. (\ref{taualphabeta}):\begin{align}
\theta^{xy}&=\theta^1,\\
\theta^{yz(zx)}&=-\frac12(\theta^1\pm\sqrt3\theta^3),\\
\bar\theta^{xy}&=\theta^3,\\
\bar\theta^{yz(zx)}&=\frac12(-\theta^3\pm\sqrt3\theta^1).
\end{align}

Let us analyze some symmetries of the Hamiltonian. In general, we can define a six-component
column vector of Majorana fermions $\zeta=(\eta^{1},\dots,\theta^{3})^{t}$ that    transform as 
\begin{equation}
\zeta'=R\zeta,
\end{equation}
where   $R$ is an SO(6) matrix.
Although the Hamiltonian in Eq. (\ref{eq:Majorana main Hamiltonian}) is not invariant
under global SO(6) transformations, it is invariant under a subset
that  includes global rotations of the form $R=R_{\eta}\oplus\mathit{I}_{\theta}$,
where $R_{\eta}$ corresponds to an SO(3) rotation of the vector $\boldsymbol{\eta}=(\eta^1,\eta^2,\eta^3)$,
and $\mathit{I}_{\theta}$ is the identity matrix in the $\theta$
sector. This symmetry is nothing but the global SU(2) invariance of Hamiltonian
(\ref{eq:main Hamiltonian}) expressed in terms of Majorana fermions. Moreover,
we can identify the $Z_{3}$ symmetry  as being generated by the transformation $R=\mathit{I}_{\eta}\oplus M_{\theta}$,
where $M_{\theta}$ is the $2\pi/3$ rotation matrix acting on the
two-component vector $(\theta^{1},\theta^{3})$ leaving $\theta^{2}$ invariant. 

The action of time reversal $T$ on Eq. (\ref{eq:Majorana main Hamiltonian})
 follows from the symmetry properties of  $\textbf{s}$ and $\boldsymbol{\tau}$ 
  discussed in   Subsection \ref{sub:ionic-gs}. In the representation of  
Eq. (\ref{eq:Majorana rep}), $T$ can be defined as complex conjugation
supplemented by $T^{-1}\theta^{2}T=-\theta^{2}$, while leaving the
other flavors invariant. With this rule, the Hamiltonian in Eq.  (\ref{eq:Majorana main Hamiltonian})
is explicitly  time-reversal invariant.

We  construct a mean-field theory with the expectation values
of bond operators $\langle\zeta_{i}^{a}\zeta_{j}^{b}\rangle$ as order
parameters. These parameters are chosen in a way that preserves as
many symmetries of the Hamiltonian (\ref{eq:Majorana main Hamiltonian})
as possible. Since the fcc lattice contains triangular plaquettes,   a Majorana QSL necessarily breaks time reversal
and reflection symmetries \cite{Fiete2012}. As a result, the mean-field
theory can preserve at most the SO(3), $Z_{3}$ and some point-group
symmetries. To preserve the SO(3) symmetry, the state must remain
invariant under any global rotation of $\textbf{s}$. Consequently,
all order parameters of the type $\langle\eta_{i}^{a}\eta_{j}^{b}\rangle$
with $a\neq b$ vanish, and $\langle\eta_{i}^{a}\eta_{j}^{a}\rangle=\langle\eta_{i}^{b}\eta_{j}^{b}\rangle$,
for $a,b=1,2,3$. Similarly, $\langle\eta_{i}^{a}\theta_{j}^{b}\rangle$
vanishes for any pair $(a,b)$. The $Z_{3}$ symmetry rotates the
$\tau^{\alpha\beta}$ operators among themselves. Thus, requiring
$Z_{3}$ invariance implies that $\langle\theta^{\alpha\beta}\theta^{2}\rangle$
must also be zero. Applying these restrictions, we  perform the mean-field
decoupling of Eq. (\ref{eq:Majorana main Hamiltonian}) and obtain
the mean-field Hamiltonian 
\begin{align}
H_{\text{MF}}  =&\frac{J}{36}\underset{\langle i,j\rangle_{\gamma}}{\sum}(3u_{ij}^{2}+3u_{ij}\bar{w}_{ij}+w_{ij}v_{ij})\nonumber \\
 & +\frac{J}{36}\underset{\langle i,j\rangle_{\gamma}}{\sum}\left[i(2u_{ij}+\bar{w}_{ij}^{\alpha\beta})\boldsymbol{\eta}_{i}\cdot\boldsymbol{\eta}_{j}\right.\nonumber \\
 & \left.+3iu_{ij}\bar{\theta}_{i}^{\alpha\beta}\bar{\theta}_{j}^{\alpha\beta}+iw_{ij}^{\alpha\beta}\theta_{i}^{2}\theta_{j}^{2}+iv_{ij}\theta_{i}^{\alpha\beta}\theta_{j}^{\alpha\beta}\right],\label{eq:Majorana MF Hamiltonian}
\end{align}
where $iu_{ij}=\langle\eta_{i}^{a}\eta_{j}^{a}\rangle$, $iv_{ij}=\langle\theta_{i}^{2}\theta_{j}^{2}\rangle$,
$iw_{ij}^{\alpha\beta}=\langle\theta_{i}^{\alpha\beta}\theta_{j}^{\alpha\beta}\rangle$
and $i\bar{w}_{ij}^{\alpha\beta}=\langle\bar{\theta}_{i}^{\alpha\beta}\bar{\theta}_{j}^{\alpha\beta}\rangle$.
Notice that all quadratic terms are diagonal in the flavor index except those involving  $\theta^{1}$ and $\theta^{3}$. Moreover, the
decoupled terms for $\eta^a$ and $\theta^{2}$ fermions  differ
only by the corresponding mean-field  amplitudes.

\begin{figure}
\begin{centering}
\subfloat[\label{fig:Ansatz choice}]{\includegraphics[clip,width=0.2\columnwidth]{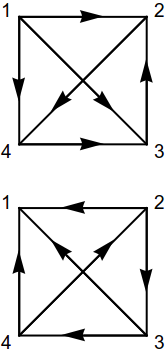}

}\hspace{.7cm}\subfloat[\label{fig:Ansatz flux}]{\includegraphics[clip,width=0.45\columnwidth]{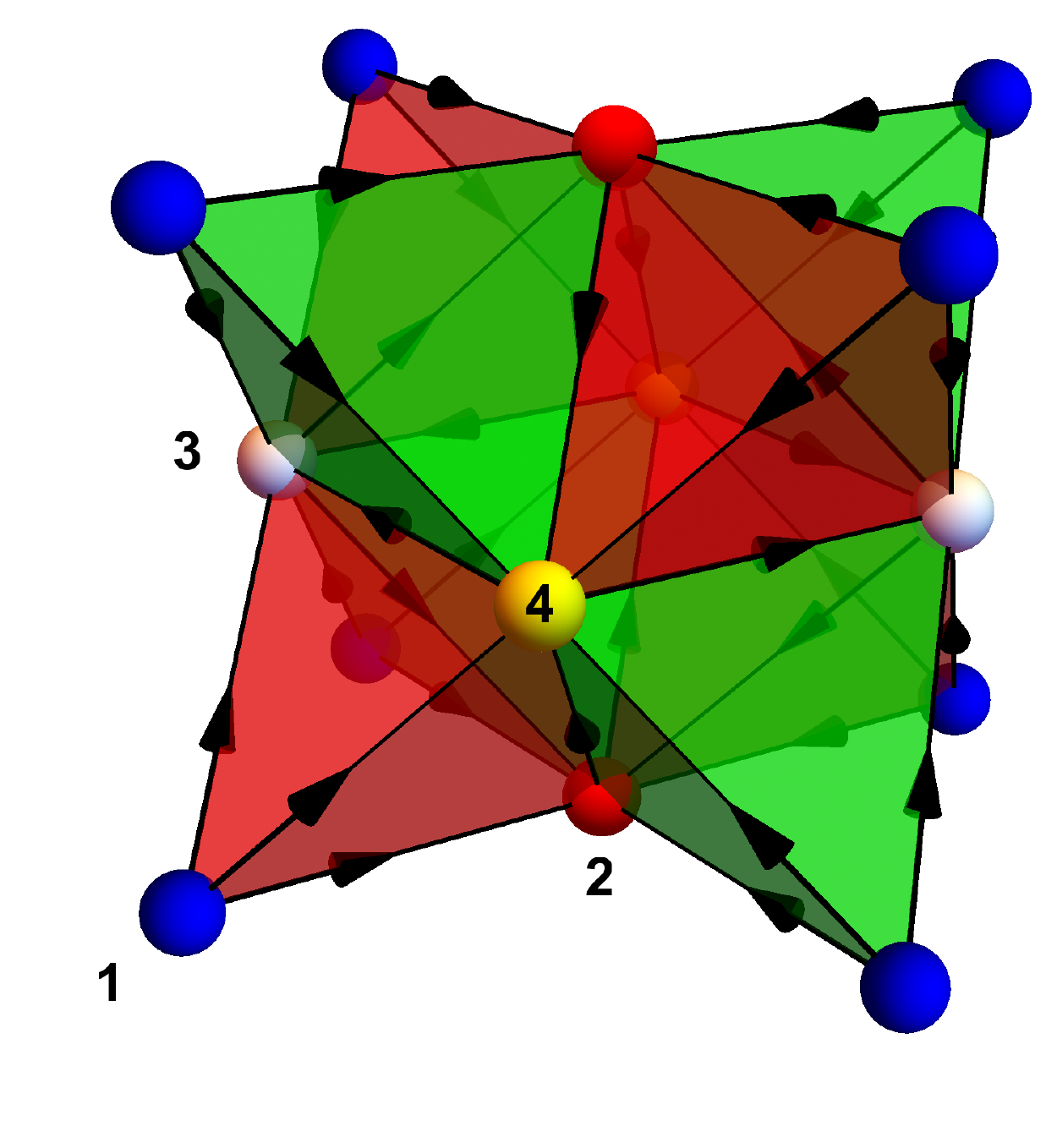} 

}
\par\end{centering}

\caption{\label{fig:ans=00003D0000E4tze}(Color online) (a) Diagrammatic representation
of the most symmetric gauge choices on an elementary tetrahedron. The arrow pointing from site $i$ to site $j$    represents that  $\phi_{ij}=+1$ (and $\phi_{ji}=-1$ in the opposite direction). Notice that the two ans\"atze are conjugated by  time reversal. (b) Representation of the $Z_{2}$
fluxes of the ansatz on the fcc lattice. The flux through each face
of a green (red) tetrahedron is positive (negative) when the sites
are oriented counterclockwise with respect to a normal vector pointing
outward. }
\end{figure}

Our study of Eq. (\ref{eq:Majorana MF Hamiltonian}) is also restricted
to translationally invariant ans\"atze. In this regard, one must
impose the magnitude of each order parameter to be uniform:
\begin{align}
u_{ij} & =u\phi_{ij}, \\
v_{ij} & =v\phi_{ij}, \\
w_{ij}^{\alpha\beta} & =w\phi_{ij}\text{\quad}\text{ for }\langle i,j\rangle_{\gamma}, \\
\bar{w}_{ij}^{\alpha\beta} & =\bar{w}\phi_{ij}\text{\quad}\text{ for }\langle i,j\rangle_{\gamma},
\end{align}
where $\phi_{ij}=\pm1$ are $Z_{2}$ link variables. The anticommutation
relations of the Majorana fermions ensure that $\phi_{ij}=-\phi_{ji}$,
which gives an orientation to the links between the sites. To orient the links, it is 
 convenient to subdivide the fcc lattice into four cubic sublattices, labeled by $X=1,2,3,4$.
As can be seen in Fig. \ref{fig:Ansatz flux}, there is a correspondence
between the sublattice sites and the vertices of   elementary
tetrahedra. The orientation of $\phi_{ij}$ between sublattices  can be represented on
a plane as shown in Fig. \ref{fig:Ansatz choice}. With two possible
values for each $\phi_{ij}$, there are in total $2^{6}=64$ different
``hopping'' configurations, which can be grouped into eight non-gauge-equivalent
ans\"atze. 

While the order parameters are not gauge invariant, physically distinct
ans\"atze  can be labeled  by  the gauge-invariant
$Z_{2}$ fluxes through the elementary plaquettes. Choosing three
nearest-neighbor sites $(i,j,k)$ in a fixed orientation, one can
define the flux $\chi_{ijk}\equiv-i\phi_{ij}\phi_{jk}\phi_{ki}$. 
The latter is closely related to  the scalar
spin chirality of $S=1/2$ QSLs \cite{Baskaran1989,Wen1989,Wen2002,Wen2004}.
Using Eq. (\ref{eq:Majorana rep}), we can write 
\begin{equation}
\textbf{s}_{i}\cdot(\textbf{s}_{j}\times\textbf{s}_{k})=-\frac{i}{8}\epsilon_{abc}\eta_{i}^{a}\eta_{j}^{a}\eta_{j}^{b}\eta_{k}^{b}\eta_{k}^{c}\eta_{i}^{c}.\label{eq:chiral}
\end{equation}
The operator in Eq. (\ref{eq:chiral}) is odd under time reversal
and is analogous to the spin chirality order parameter. Using our mean-field
decoupling, we obtain 
\begin{equation}
\left\langle \textbf{s}_{i}\cdot(\textbf{s}_{j}\times\textbf{s}_{k})\right\rangle =\frac{3}{8}u_{ij}u_{jk}u_{ki}=\frac{3}{8}iu^{3} \chi_{ijk} .\label{eq:chiral parameter}
\end{equation}
The physical state is determined by the $Z_{2}$ flux configuration
on all plaquettes. We should note that 
the fluxes through the faces of any tetrahedron are not all independent. If the sites on any
given face are oriented clockwise with respect to an outward normal
vector, the four fluxes obey the relation $\prod_{r=1}^{4}\chi_{r}=1$,
where $r$ labels the faces of the tetrahedron.

Time reversal plays an important role in choosing the mean-field theory,
since it relates pairs of non-equivalent gauge configurations (see
Fig. \ref{fig:Ansatz choice}). In terms of $Z_{2}$ fluxes, $T$
inverts $\chi_{ijk}$ of every elementary plaquette of the lattice.
Although not related by gauge transformations, two gauge choices related
by $T$ lead to degenerate mean-field ground states. Still guided
by symmetry principles, we study here the most symmetric ans\"atze,
which are characterized by the same $Z_{2}$ flux through all faces
of a tetrahedron. The imposition of translation invariance implies
that two tetrahedra sharing an edge have opposite $Z_{2}$ fluxes
(see Fig. \ref{fig:Ansatz flux}). In other words, our ansatz  is a staggered-flux Majorana QSL, where the staggering is
between nearest-neighbor tetrahedra.

We define the parity transformation $P$ as a reflection by a symmetry plane of the fcc lattice. As can be seen in Fig. \ref{fig:Ansatz flux},  $P$ inverts all flux orientations in the mean-field ansatz.
Since the Majorana QSL breaks both $P$ and $T$ symmetries, it is classified as a chiral spin(-orbital) liquid \cite{Wen1989}. However,
notice that the antiunitary operator $PT$ is still a symmetry, with
$(PT)^{2}=+1$.  
Since we are dealing with  system of fractionalized quasiparticles,
  crystalline symmetries must be studied
by means of  a projective symmetry group (PSG) analysis \cite{Wen2004}, which
was discussed in Ref. \cite{Natori2016}. Due to the breaking of $P$ and $T$, the point group symmetry of   Hamiltonian
(\ref{eq:Majorana main Hamiltonian}) is reduced from $O_{h}\times Z_{2}$,
with $Z_{2}$ corresponding to time reversal, to a group isomorphic
to $O_{h}$. 

After fixing the ansatz, we solve the mean-field Hamiltonian using
the Fourier mode expansion 
\begin{equation}
\zeta_{\textbf{k}X}^{a}=\sqrt{\frac{2}{N}}\underset{j\in X}{\sum}\zeta_{jX}^{a}e^{-i\textbf{k}\cdot\textbf{R}_{j}},\label{eq:majorana k}
\end{equation}
where $X=1,2,3,4$ is the sublattice  index and $N$ is the total number
of sites in the fcc lattice. The positions of the sites in sublattice
$X$ are given by
\begin{equation}
\textbf{R}_{j}=(n_{x},n_{y},n_{z})+\boldsymbol{\delta}_{X},\quad n_{a}\in\mathbb{Z},
\end{equation}
with $\boldsymbol{\delta}_{1}=(0,0,0)$, $\boldsymbol{\delta}_{2}=(1/2,1/2,0)$,
$\boldsymbol{\delta}_{3}=(0,1/2,1/2)$, and $\boldsymbol{\delta}_{4}=(1/2,0,1/2)$ in units where the  lattice parameter is set to 1.
The operators  $\zeta_{\textbf{k}X}^{a}$ obey $(\zeta_{\textbf{k}X}^{a})^\dagger=\zeta_{-\textbf{k}X}^{a}$ and $\{\zeta_{\textbf{k}X}^{a},\zeta_{\textbf{k}'X'}^{a}\}=\delta_{\textbf{k},-\textbf{k}'}\delta_{X,X^{\prime}}$.
Thus,  $\zeta_{\textbf{k}X}^{a}$ can then be treated as complex fermions with
well-defined occupation numbers   if we split  the first Brillouin zone
of the cubic lattice into two halves, which can be mapped into each
other by inversion. Only one of these halves is taken into account
and will be called $\frac{1}{2}\text{BZ}$. It is worth pointing out that
the PSG analysis shows that the mean-field ansatz
is invariant under translations on the fcc lattice \cite{Natori2016}. This can be understood
intuitively by noting that translations by $\boldsymbol{\delta}_{X}$
exchange the sublattices but do not change the signs of the gauge-invariant
fluxes represented in Fig. \ref{fig:Ansatz flux}.

The mean-field Hamiltonian in Eq. (\ref{eq:Majorana MF Hamiltonian})
can be rewritten in the form 
\begin{align}
H_{\text{MF}}  =&\frac{NJ}{2}\left(u^{2}+u\bar{w}+\frac{vw}{3}\right)\nonumber \\
 & +\frac{J}{18}\underset{\textbf{k}\in\frac{1}{2}BZ}{\sum}\left[(2u+\bar{w})\underset{a=1}{\overset{3}{\sum}}\left(\eta_{\textbf{k}}^{a}\right)^{\dagger}\mathcal{H}_{1}(\textbf{k})\eta_{\textbf{k}}^{a}\right.\nonumber \\
 & \left.+w\left(\theta_{\textbf{k}}^{2}\right)^{\dagger}\mathcal{H}_{1}(\textbf{k})\theta_{\textbf{k}}^{2}+\left(\Theta_{\textbf{k}}\right)^{\dagger}\mathcal{H}_{2}(\textbf{k})\Theta_{\textbf{k}}\right],\label{eq:MatrixHmf}
\end{align}
where $\zeta_{\mathbf{k}}=(\zeta_{\mathbf{k},1},\zeta_{\mathbf{k},2},\zeta_{\mathbf{k},3},\zeta_{\mathbf{k},4})^{t}$ 
for  $\zeta\in\{\eta^{a},\theta^{2}\}$ are four-component spinors, and $\Theta_{\textbf{k}}=(\theta_{\textbf{k}1}^{1},\theta_{\textbf{k}2}^{1},...,\theta_{\textbf{k}4}^{3})^{t}$ is an eight-component spinor.
To find the ground state of Eq. (\ref{eq:MatrixHmf}),   first we
study the $4\times4$ matrix $\mathcal{H}_{1}(\textbf{k})$,   given by 
\begin{equation}
\mathcal{H}_{1}(\textbf{k})=\textbf{h}(\textbf{k})\cdot\boldsymbol{\Sigma},\label{H1kmatrix}
\end{equation}
with 
\begin{align}
\textbf{h}(\textbf{k}) &   =(h_{1}(\textbf{k}),h_{2}(\textbf{k}),h_{3}(\textbf{k}))\nonumber\\
&=4\left(\cos\frac{k_{x}}{2}\cos\frac{k_{y}}{2},\cos\frac{k_{y}}{2}\cos\frac{k_{z}}{2},\cos\frac{k_{x}}{2}\cos\frac{k_{z}}{2}\right),
\end{align}
and 
\begin{align}
\boldsymbol{\Sigma}&= (\Sigma_{1},\Sigma_{2},\Sigma_{3})\nonumber\\
&=(-\sigma^z\otimes\sigma^y,-\sigma^y\otimes \mathbb I,-\sigma^x\otimes\sigma^y).
\end{align}
Let $U_{\textbf{k}}$ be the unitary matrix that diagonalizes $\mc H_1(\mathbf k)$:
\begin{equation}
U_{\textbf{k}}^{\dagger}\mathcal{H}_{1}(\textbf{k})U_{\textbf{k}}=\Lambda_{1}(\textbf{k}),\label{eq:defUk}
\end{equation}
where $\Lambda_{1}(\textbf{k})$ is   diagonal.
Since the matrices in Eq. (\ref{H1kmatrix}) obey the Clifford algebra $\{\Sigma^{a},\Sigma^{b}\}=2\delta^{ab}$,
the eigenvalues of $\mathcal{H}_{1}(\mathbf{k})$ are simply $\pm|\mathbf{h}(\mathbf{k})|$
and are doubly degenerate. This is a Kramers-type degeneracy that
can be explained by point group symmetries, as discussed in Ref. \cite{Natori2016}.
The mean-field Hamiltonian is diagonal in the basis of operators  $\tilde{\zeta}_{\textbf{k}\lambda}$ given
by 
\begin{equation}
\zeta_{\textbf{k}X}=\underset{\lambda=1}{\overset{4}{\sum}}\left(U_{\textbf{k}}\right)_{X\lambda}\tilde{\zeta}_{\textbf{k}\lambda},\label{eq:Uk eigen}
\end{equation}
with $\lambda=1,\dots,4$ being the band index. The order parameters
$u$ and $v$ are determined by self-consistent equations:
\begin{align}
u  =&-i\left\langle \eta_{j,1}^{1}\eta_{j+\boldsymbol{\delta}_{2},2}^{1}\right\rangle \nonumber \\
 =&\frac{16}{N}\text{Im} \underset{\textbf{k}}{\sum}e^{i\textbf{k}\cdot\boldsymbol{\delta}_{2}}\underset{\lambda}{\overset{}{\sum}}\left(U_{\textbf{k}}\right)_{2\lambda} (U_{\textbf{k}}^{\dagger} )_{\lambda1} \langle \left(\tilde{\eta}_{\textbf{k}\lambda}^{1}\right)^{\dagger}\tilde{\eta}_{\textbf{k}\lambda}^{1} \rangle  ,\label{eq:self-consistent parameters 1}\\
v  =&-i\left\langle \theta_{j,1}^{2}\theta_{j+\boldsymbol{\delta}_{2},2}^{2}\right\rangle \nonumber \\
  =&\frac{16}{N}\text{Im}\underset{\textbf{k}}{\sum}e^{i\textbf{k}\cdot\boldsymbol{\delta}_{2}}\underset{\lambda}{\overset{}{\sum}}\left(U_{\textbf{k}}\right)_{2\lambda}(U_{\textbf{k}}^{\dagger})_{\lambda1}\langle (\tilde{\theta}_{\textbf{k}\lambda}^{2})^{\dagger}\tilde{\theta}_{\textbf{k}\lambda}^{2}\rangle ,\label{eq:self-consistent parameters}
\end{align}
 where the sum over $\mathbf k$ is restricted to $\mathbf k\in\frac{1}{2}\text{BZ}$.  At zero temperature, we can replace the average occupation of the single-particle states by 
\bea
\langle(\tilde{\eta}_{\textbf{k}\lambda}^{a})^\dagger\tilde{\eta}^a_{\textbf{k}\lambda}\rangle&=&\Theta(-\epsilon_{\textbf{k}\lambda}^{(\eta)}),\label{eq:Heaviside}\\
\langle(\tilde{\theta}_{\textbf{k}\lambda}^{2})^\dagger\tilde{\theta}^2_{\textbf{k}\lambda}\rangle&=&\Theta(-\epsilon_{\textbf{k}\lambda}^{(\theta^2)}),\label{eq:Heaviside2}
\eea
where $\Theta(x)$ is the Heaviside step function and  \bea
\epsilon_{\textbf{k}\lambda}^{(\eta)}&=&\frac{J(2u+\bar w)}{18}|\mathbf h(\mathbf k)| C_\lambda,\label{eq:eta dispersion}\\
\epsilon_{\textbf{k}\lambda}^{(\theta^2)}&=&\frac{Jw}{18} |\mathbf h(\mathbf k)| C_\lambda\label{eq:th2 dispersion}
\eea
are the dispersion relations of the $\eta$ and $\theta^2$ fermions, respectively, with $C_\lambda=-1$ for $\lambda=1,2$ and $C_\lambda=+1$ for $\lambda=3,4$.

The expressions
for $u$ and $v$ coincide at zero temperature, except possibly for
a sign depending on the relative sign between the parameters $2u+\bar{w}$
and $w$. Without loss of generality,
we fix $u>0$ (which corresponds to fixing the sign of the $T$-symmetry-breaking
order parameter). As discussed in Ref. \cite{Natori2016}, the two
cases $v=u$ or $v=-u$ give rise to two different ans\"atze, with different
expressions for the $8\times8$ matrix $\mathcal{H}_{2}(\mathbf{k})$.
In the remainder of this work, we will deal with the case $v=u$ \cite{comentario}.
In this case, self-consistency of the mean-field equations implies
 $2u+\bar{w}>0$ and $w>0$ .

Having fixed   $\text{sgn}(uv)>0$, we find that the $8\times8$
matrix $\mathcal{H}_{2}(\textbf{k})$ in Eq. (\ref{eq:MatrixHmf}) is given by 
\begin{align}
\mathcal{H}_{2}(\textbf{k}) & =\left(\begin{array}{cc}
\mathcal{H}_{\theta_{1}\theta_{1}}(\textbf{k}) & \mathcal{H}_{\theta_{1}\theta_{3}}(\textbf{k})\\
\mathcal{H}_{\theta_{1}\theta_{3}}(\textbf{k}) & \mathcal{H}_{\theta_{3}\theta_{3}}(\textbf{k})
\end{array}\right),
\end{align}
where 
\begin{align}
\mathcal{H}_{\theta_{1}\theta_{1}}(\mathbf{k}) & =v\text{}h_{1}(\mathbf{k})\Sigma_{1}+\frac{9u+v}{4}\left[h_{2}(\mathbf{k})\Sigma_{2}+h_{3}(\mathbf{k})\Sigma_{3}\right],\nonumber \\
\mathcal{H}_{\theta_{3}\theta_{3}}(\mathbf{k}) & =3u\text{}h_{1}(\mathbf{k})\Sigma_{1}+\frac{3}{4}(u+v)\left[h_{2}(\mathbf{k})\Sigma_{2}+h_{3}(\mathbf{k})\Sigma_{3}\right],\nonumber \\
\mathcal{H}_{\theta_{1}\theta_{3}}(\mathbf{k}) & =\frac{\sqrt{3}}{4}(3u-v)\left[-h_{2}(\mathbf{k})\Sigma_{2}+h_{3}(\mathbf{k})\Sigma_{3}\right].\label{eq:H2}
\end{align}
We denote by $V_{\mathbf{k}}$ the matrix  that diagonalizes $\mathcal{H}_{2}(\mathbf{k})$:
\begin{equation}
V_{\mathbf{k}}^{\dagger}\mathcal{H}_{2}(\mathbf{k})V_{\mathbf{k}}=\Lambda_{2}(\mathbf{k}).\label{eq:defVk}
\end{equation}
The order parameters $w$ and $\bar{w}$ can be  calculated similarly
to Eq. (\ref{eq:self-consistent parameters}), using the components
of $V_{\mathbf{k}}$ instead of $U_{\mathbf{k}}$.

\begin{figure}
\begin{centering}
\subfloat[\label{fig:Fermi lines}]{\includegraphics[clip,width=0.3\columnwidth]{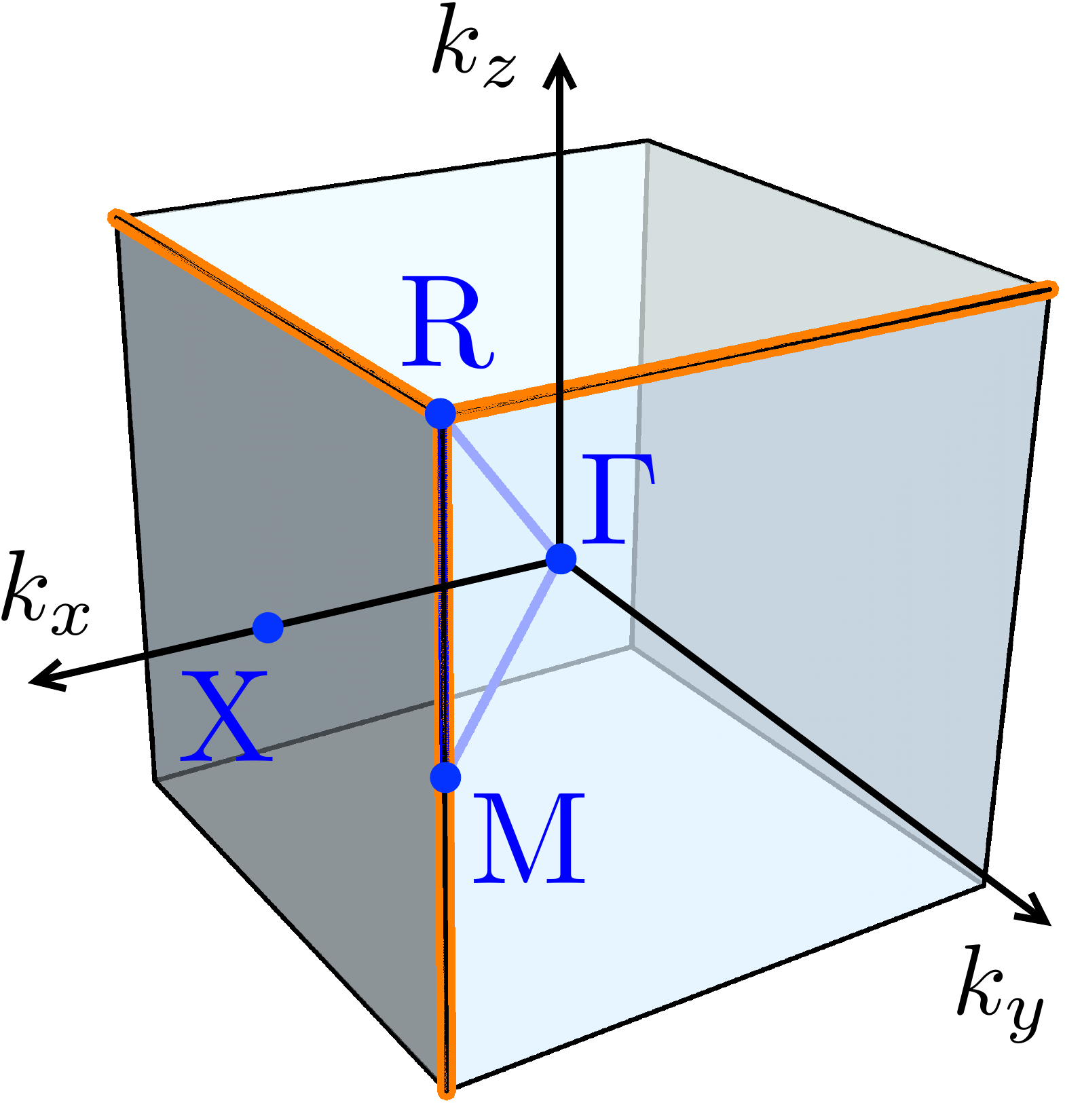} 

}\hspace{.2cm}\subfloat[\label{fig:Dispersion fermions}]{\includegraphics[clip,width=0.63\columnwidth]{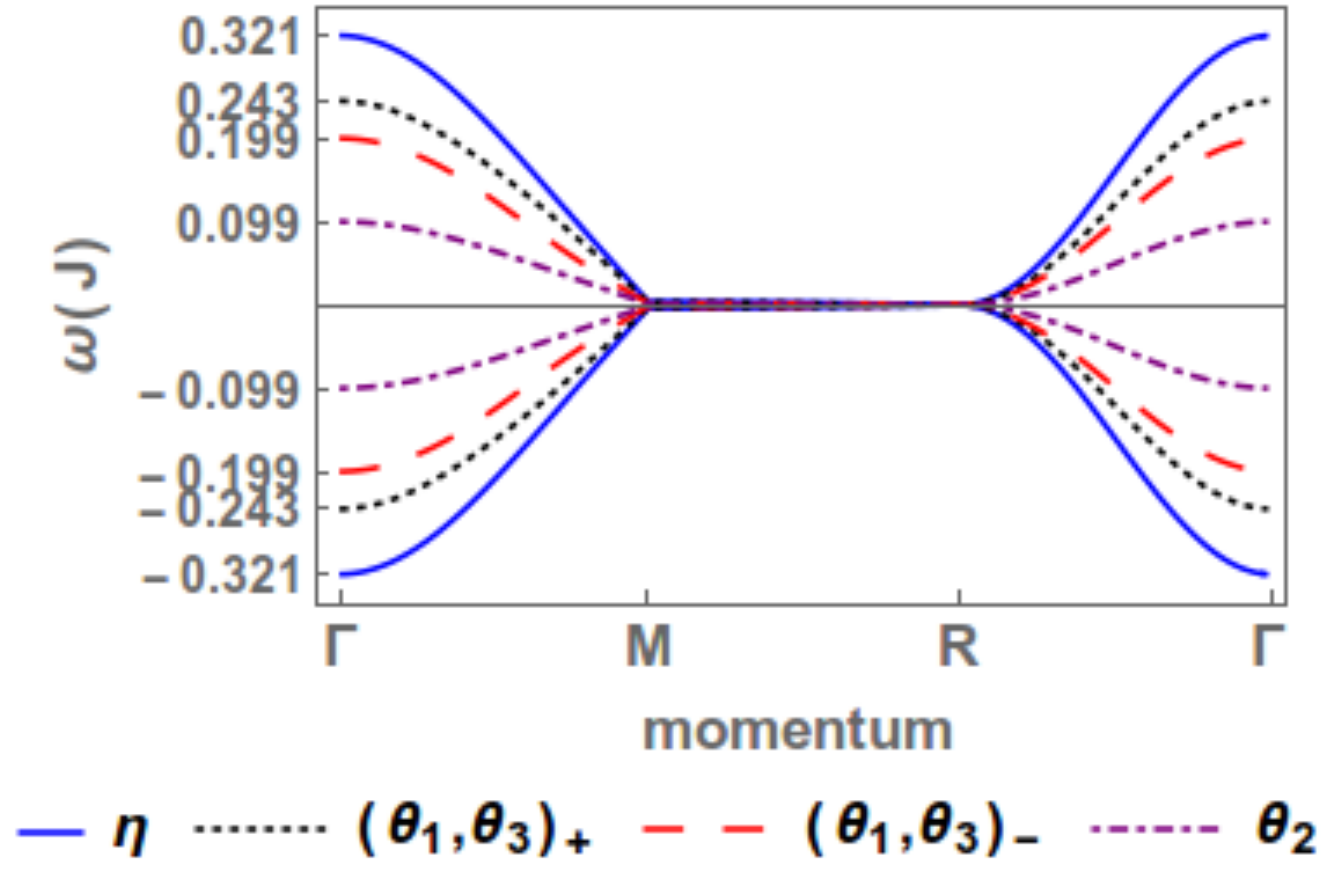} 

}
\par\end{centering}

\caption{\label{fig:band}(Color online) (a) First Brillouin zone of the cubic
lattice highlighting the Fermi lines (orange lines). (b) Dispersion
for different fermion flavors.}
\end{figure}

Figure \ref{fig:band} shows the dispersion relation for the different
flavors of Majorana fermions. The bands are particle-hole symmetric
and doubly degenerate for all flavors. The dispersion relations  of all bands
are qualitatively similar, differing mainly by their bandwidths. The
most remarkable feature is that the band structure displays nodal
lines along the edges of the Brillouin zone, a consequence of the
vanishing of $\mathbf{h}(\mathbf{k})$ when two components of $\mathbf{k}$
are equal to $\pi$ \cite{Natori2016}. The energy increases linearly with the distance
in momentum space from a generic point on a nodal line. The exception
is the vertex point $R=(\pi,\pi,\pi)$, where the nodal lines cross
and the dispersion becomes approximately quadratic but anisotropic.
For $\mathbf{k}=(\pi,\pi,\pi)+\mathbf{q}$, with $|\mathbf{q}|\ll1$,
we obtain for all bands
\begin{equation}
\epsilon_{\mathbf{k}\lambda}\propto\sqrt{q_{x}^{2}q_{y}^{2}+q_{y}^{2}q_{z}^{2}+q_{z}^{2}q_{x}^{2}},\label{eq:quadratic dispersion}
\end{equation}
 which is of the form $\epsilon_{\mathbf{k}\lambda}=q^{2}f_{\lambda}(\Omega)$,
with $f_{\lambda}(\Omega)$ a function of the spherical angle coordinates
of $\mathbf{q}$. 

The single-particle states in the neighborhood of the $R$ point dominate the low-energy
physics due to the quadratic dispersion.
To see this, we can compute the corresponding contribution to the density of states 
\begin{align}
\rho_{\text{point}}(E) & =\underset{\textbf{k},\lambda}{\sum}\delta(E-\epsilon_{\mathbf{k}\lambda})\nonumber \\
 & \approx\sum_{\lambda}\int d\Omega\int\frac{dq\,q^{2}}{(2\pi)^{3}}\,\delta(E-q^{2}f_{\lambda}(\Omega))\nonumber \\
 & =\frac{1}{2}\sqrt{E}\sum_{\lambda}\int d\Omega[f_{\lambda}(\Omega)]^{-3/2}.\label{eq:density states}
\end{align}
Thus, we find  $\rho_{\text{point}}(E)\propto\sqrt{E}$, a vanishing
density of states characteristic of a pseudogap. The same analysis
for the density of states around generic points on the nodal line
parallel to the $k_{z}$ axis yields
\begin{align}
\rho_{\text{line}}(E) & \approx\underset{\lambda}{\sum}\int dk_{z}\int d\varphi\int\frac{dp\text{\,}p}{(2\pi)^{3}}\delta(E-v_{\lambda}(k_{z})p)\nonumber \\
 & =E\underset{\lambda}{\sum}\int_{0}^{\pi-\epsilon}\frac{dk_{z}}{4\pi^{2}[v_{\lambda}(k_{z})]^{2}},\label{DOSline}
\end{align}
where $v_{\lambda}(k_{z})$ is the effective velocity of the linear
dispersion around the nodal line and we cut off the integral at $|k_{z}-\pi|=\epsilon>0$
to exclude the contribution from the $R$ point {[}since $v_{\lambda}(k_{z}\to\pi)\to0${]}.
Thus, the contribution from the nodal lines to the density of states
is $\rho_{\text{line}}(E)\propto E$. This  is the same result as for a Dirac point
in two dimensions. The comparison of Eqs. (\ref{eq:density states}) and (\ref{DOSline}) suggests that  the low-temperature thermodynamics
of the chiral spin-orbital liquid should be governed by the quadratic
band touching point.

\section{Specific heat, spin-lattice relaxation rate and dynamical spin structure
factor \label{sec:usual probes}}

In the absence of a ``smoking-gun'' signature of QSLs \cite{Savary2016},
a proper characterization of such states must combine the response
to different perturbations. In this section, we calculate the response
of our chiral spin-orbital liquid to three well-established probes:
specific heat, nuclear magnetic resonance, and inelastic neutron scattering.

\subsection{Specific Heat\label{sec:CV}}

The specific heat $C_{V}$ of Ba$_{2}$YMoO$_{6}$ was measured by
de Vrie \textit{et al.} \cite{deVries2010} and Aharen \textit{et
al.} \cite{Aharen2010}. In both experiments, the magnetic contribution
was obtained by subtracting off the data for the isostructural nonmagnetic
compound Ba$_{2}$YNbO$_{6}$ from the total specific heat of Ba$_{2}$YMoO$_{6}$.
The measurements agree about the presence of a broad peak around 50
K. However, the reported values of $C_{V}$ at the maximum are different:
$7.5$ J/mol$\cdot$K in Ref. \cite{deVries2010} versus $2.5$ J/mol$\cdot$K
in Ref. \cite{Aharen2010}. By integrating $C_{V}$ out to $T\approx200$
K, de Vries et al. \cite{deVries2010} found that the entropy released
is close to $k_{B}\ln4$, as expected for a $j=3/2$ system. In Ref.
\cite{deVries2013}, the low-temperature behavior of $C_{V}$ was
interpreted as evidence for a pseudogap in magnetic excitations. On
the other hand, Aharen \textit{et al.} \cite{Aharen2010} noted that
the entropy lost below 50 K is lower than $k_{B}\ln2$ and found an
abrupt drop in the magnetic specific heat above 60 K. While it would
be desirable to clarify the disagreement between these experiments,
here we will focus on the common observation of a broad peak in $C_{V}$
and use this information to set the energy scale in our spin-orbital
model.

To calculate $C_{V}$, we follow the method of Ref. \cite{Herfurth2013},
which studied a Majorana QSL on a $S=1/2$ Heisenberg model. The mean-field
theory described in Sec. \ref{sec:QSL-MFT} can be extended to $T>0$
by replacing the average occupation of single-particle states by the Fermi-Dirac
distribution:
\begin{align}
\langle\tilde{\zeta}_{\mathbf{k}\lambda}^{\dagger}\tilde{\zeta}^{\phantom\dagger}_{\mathbf{k}\lambda}\rangle & =n_{F}(\epsilon^{(\zeta)}_{\mathbf{k}\lambda})\nonumber \\
 & =\left[1+\exp(\beta\epsilon^{(\zeta)}_{\mathbf{k}\lambda})\right]^{-1}\label{eq:T>0 parameter}
\end{align}
where $\beta=1/(k_{B}T)$. The order parameters calculated using Eq.
(\ref{eq:T>0 parameter}) define a temperature-dependent mean-field
Hamiltonian $H_{\text{MF}}(T)$. We fix these parameters by minimizing
the free energy, 
\begin{equation}
F=-\frac{1}{\beta}\underset{\textbf{k}\in\frac{1}{2}BZ}{\sum}\underset{\lambda}{\sum}\ln(1+e^{-\beta\epsilon_{\textbf{k}\lambda}})+\frac{NJ}{2}\left(u^{2}+u\bar{w}+\frac{vw}{3}\right),\label{eq:Gen formula free energy}
\end{equation}
and solving the self-consistent equations numerically. The absolute
values of the order parameters decrease with increasing temperature
as shown in Fig. \ref{fig:order parameters}.

We analyze the free energy for small values of the order parameters
in Appendix \ref{sec:Landau's free energy} and show that they vanish
at the critical temperature $k_{B}T_{c}=J/12$. The parameters $u$
and $\bar{w}$ vanish as $(T_{c}-T)^{1/2}$, as expected for primary
order parameters at the mean-field level. Note that this is a well-defined
second-order phase transition because a nonzero value of $u$ implies
spontaneous breaking of time reversal symmetry {[}see Eq. (\ref{eq:chiral parameter}){]}.
On the other hand, $v$ and $w$ behave as secondary order parameters
\cite{Stokes1991,Hatch2001} and vanish as $(T_{c}-T)^{3/2}$ (see
Fig. \ref{fig:order parameters}).

At low temperatures $T\ll J$, we can approximate the order parameters
by their zero-temperature values. The main effect of thermal fluctuations
in this regime is to change the occupation of the states in a band
with fixed bandwidth. Using the density of states in Eq. (\ref{eq:density states}),
we find 
\begin{align}
C_{V}(T\ll J) & =\int_{0}^{\infty}dE\,E\rho_{\text{point}}(E)\frac{\partial n_{F}}{\partial T}\nonumber \\
 & \propto T^{3/2}\int_{0}^{\infty}dx\,\frac{x^{5/2}e^{x}}{(1+e^{x})^{2}}.
\end{align}
Thus, at sufficiently low temperatures we obtain the power-law behavior
$C_{V}\propto T^{3/2}$.

\begin{figure}
\begin{centering}
\subfloat[\label{fig:order parameters}]{\includegraphics[clip,width=0.8\columnwidth]{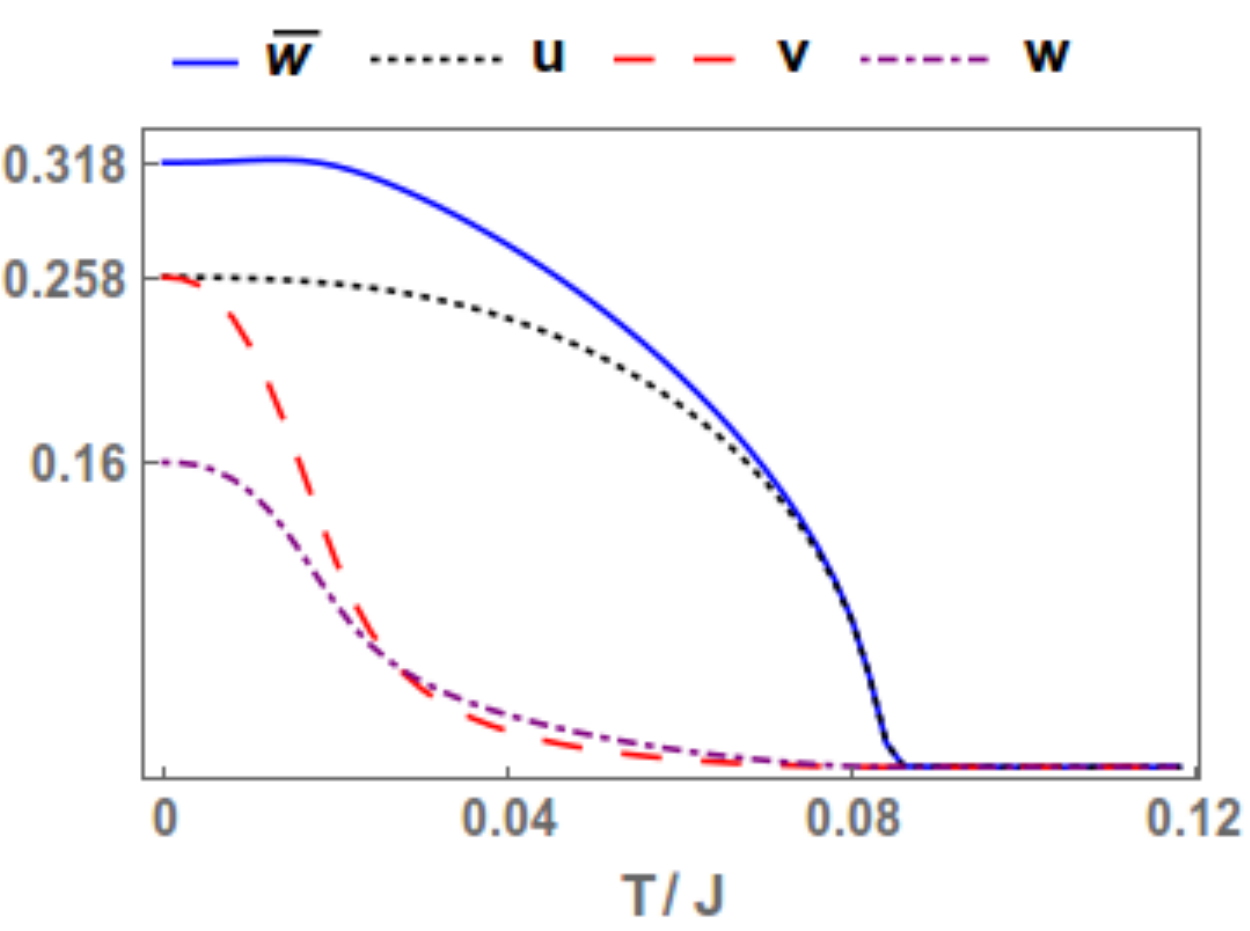}

}
\par\end{centering}

\begin{centering}
\subfloat[\label{fig:specific heat}]{\includegraphics[clip,width=0.8\columnwidth]{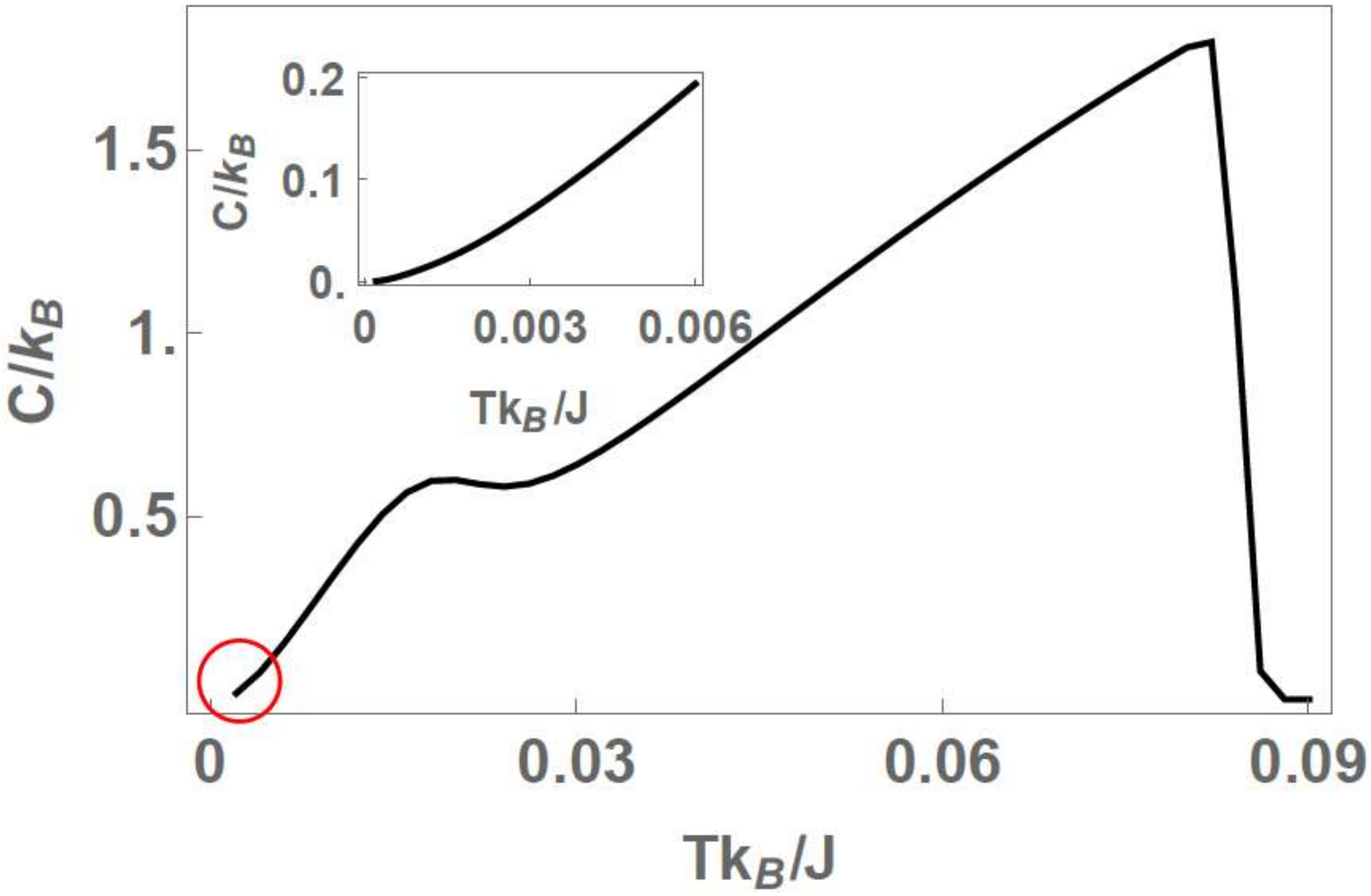} 

}
\par\end{centering}

\caption{\label{fig:Magnetic-specific-heat}(a) Absolute value of the order
parameters of the chiral spin-orbital liquid as a function of temperature.
(b) Magnetic specific heat per site calculated within the mean-field
theory.}
\end{figure}

The whole temperature dependence of the specific heat    is shown
in Fig. \ref{fig:specific heat}. Starting from the low-temperature
limit, we see that the $T^{3/2}$ behavior turns into a small plateau
at $k_{B}T\approx0.02J$. Above this temperature, there is a regime
where $C_{V}$ increases approximately linear with $T$, followed
by a  sharp drop  at $k_{B}T_{c}=J/12$ (which is a discontinuity
at the mean-field level). Our theoretical result shows qualitative
agreement with the experimental data obtained in Ref. \cite{Aharen2010}.
To make some quantitative predictions, we use the experimental data
from Ref. \cite{Aharen2010} to estimate $T_{c}\approx70$ K. This
fixes the exchange coupling constant at $J\approx72\text{ meV}$. 

The lost entropy per site calculated within the parton mean-field
theory is approximately $1.98k_{B}$, significantly higher than the
expected for a $j=3/2$ system ($k_{B}\ln4\approx1.39k_{B}$). We
expect the mean-field result to overestimate the entropy since this
approximation violates the local constraint in Eq. (\ref{eq:Majorana constraint}).
As a result, the number of microstates in this approach is higher
than the actual number of physical states. At zero temperature, this
problem was circumvented by using the Variational Monte Carlo (VMC)
method to find a better estimate of the ground state energy \cite{Natori2016}.
To our knowledge, the only calculations of thermodynamic quantities
in controlled approximations of QSLs at finite temperatures were done
recently for the Kitaev model \cite{Nasu2014,Nasu2015,Nasu2016,Yoshitake2016}.
The numerical methods benefit from the exact solvability of the Kitaev
model, a feature not available in our case.

Another problem with the mean-field approach used in this section
is that it implicitly assumes that the Z$_{2}$ gauge configuration
in the mean-field ansatz remains frozen at finite temperatures. Without
this assumption, we would not be able to diagonalize a free-fermion
Hamiltonian and find the dispersion relations  
used in Eq. (\ref{eq:T>0 parameter}). Remarkably,  studies
of thermodynamics of the Kitaev model \cite{Nasu2014,Nasu2015,Nasu2016,Yoshitake2016}
 found that thermal fluctuations of the $Z_{2}$ gauge field
are activated at temperatures \emph{much lower} than the bandwidth
of the Majorana fermions at zero temperature. The proliferation of
thermally excited visons is detected as an additional peak in the
specific heat. In the case of three-dimensional QSLs \cite{Nasu2014},
the lower-temperature peak in $C_{V}$ is a true singularity and signals
a topological phase transition predicted by $Z_{2}$ gauge theory
\cite{POLYAKOV1978477,SenthilPRB2000}.  

Nevertheless, we still argue in favor of using the
broad peak at higher temperature to determine the energy scale of
the exchange interactions. We note that in controlled numerical calculations for Kitaev models 
the corresponding  peak  in $C_V$ is   well described by the approximations of either
fixing a uniform configuration or treating the $Z_{2}$ gauge field
as a completely random variable \cite{Nasu2015}.  
In the following we will use the
estimate $J\approx72\text{ meV}$ to analyze the energy scales that
appear in INS and RIXS.

\subsection{Spin-lattice relaxation rate}

Nuclear magnetic resonance is a technique that relies on nuclear spins
  to probe the local environment. In 
spin systems, the energy transfer between electrons and nuclei is
mediated by the hyperfine coupling \begin{align}
H_{\text{hf}}= & -\textbf{I}_{i}\cdot\textbf{B}_{\text{hf}}(i),
\end{align}
where $\mathbf I_i$ is the nuclear spin at site $i$ and  $\textbf{B}_{\text{hf}}(i)$ is the hyperfine effective field due to neighboring electrons. In
the experiment of Ref. \cite{Aharen2010}, the excited nuclear spin
was the $I=1/2$ $^{89}$Y, which couples to the $j=3/2$ magnetic moment
of Mo electronic spins. We can then write 
\begin{equation}
\textbf{B}_{\text{hf}}(i)=A_0\underset{\boldsymbol{\delta}}{\sum}\textbf{J}_{i+\boldsymbol{\delta}},\label{eq:Bhf}
\end{equation}
where $\boldsymbol{\delta}$ is the relative position of the atoms
of $^{89}$Y and their neighboring Mo atoms, and $A_0$ is the constant hyperfine coupling 
for first-neighbor $\boldsymbol{\delta}$. The spin-lattice relaxation
rate $1/T_{1}$ is given by 
\begin{equation}
\frac{1}{T_{1}}\propto\frac{1}{1-e^{-\beta\omega}}\underset{\textbf{q}\in BZ}{\sum}|A(\textbf{q})|^{2}\chi_{+-}^{\prime\prime}(\textbf{q},\omega),\label{eq:1/T1}
\end{equation}
where $\omega$ is the resonance frequency, 
\begin{equation}
A(\textbf{q})=A_0\left[\cos\left(\frac{q_{x}}{2}\right)+\cos\left(\frac{q_{y}}{2}\right)+\cos\left(\frac{q_{z}}{2}\right)\right]
\end{equation}
is the hyperfine interaction form factor, 
and $\chi_{+-}^{\prime\prime}(\textbf{q},\omega)$ is the spectral
function given by 
\begin{align}
\chi_{+-}^{\prime\prime}(\textbf{q},\omega)  =&\frac{1-e^{-\beta\omega}}{Z}\underset{n,n^{\prime}}{\sum}e^{-\beta E_{n}}\left|\left\langle n^{\prime}\left|J_{\textbf{q}}^{-}\right|n\right\rangle \right|^{2}\nonumber \\
 & \times\delta(\omega-E_{n^{\prime}}+E_{n}),\label{spectralchi}
\end{align}
with $|n\rangle$ being an exact eigenstate of the spin Hamiltonian
with energy $E_{n}$, and $Z=\sum_ne^{-\beta E_n}$ being  the partition function. Here,  $J_{j}^{-}=J_{j}^{x}-iJ_{j}^{y}$ is the angular momentum
lowering operator at site $j$ and $J_{\mathbf{q}}^{-}$ is its Fourier
transform.

We    calculate $1/T_{1}$ for the chiral spin-orbital
liquid using the parton mean-field theory.  The main idea is to write $J_{j}^{-}$ in terms of $\mathbf{s}$ and
$\boldsymbol{\tau}$ according to  Table \ref{tab:multipoles} and relate the spectral function $\chi^{\prime\prime}_{+-}(\mathbf q,\omega)$ to finite-temperature correlations of free Majorana fermions. In this approach, we employ the order parameters calculated self-consistently at finite temperature as described in Subsection \ref{sec:CV}.  

To gain some insight into the low-temperature behavior of $1/T_1$, we find it instructive to first analyze the contribution of the $\eta$ fermions to the total spectral function, since in this case we can derive some closed-form expressions. Using the procedure outlined in Appendix \ref{sec:Appendix}, we find that the $\eta$-fermion contribution    in the experimentally relevant regime
$\omega\ll k_{B}T$ is given by\begin{align}
\left(\frac{1}{T_{1}}\right)_{\eta}  \propto&\int_{\text{BZ}}d^{3}kd^{3}k'\frac{|A(\mathbf{k}-\mathbf{k}')|^{2}\mathscr{F}^{\eta}(\mathbf{k},\mathbf{k}')}{\cosh^{2}\left(\beta\epsilon_{\mathbf{k}1}^{(\eta)}/2\right)}\nonumber \\
 & \times\delta(\epsilon_{\mathbf{k}'1}^{(\eta)}-\epsilon_{\mathbf{k}1}^{(\eta)}),\label{T1eta}
\end{align}
where 
\begin{equation}
\mathscr{F}^{\eta}(\mathbf{k},\mathbf{k}')=1+\frac{\mathbf{h}(\mathbf{k})\cdot\mathbf{h}(\mathbf{k}')}{|\mathbf{h}(\mathbf{k})||\mathbf{h}(\mathbf{k}')|}.\label{eq:NMR form factor}
\end{equation}

At low temperatures $k_{B}T\ll J$, the spin-lattice relaxation rate is dominated by 
 excitations with small momentum transfer  near    
the quadratic band touching  point.   We write $\textbf{k}=(\pi,\pi,\pi)+\textbf{q}$ and
$\textbf{k}'=(\pi,\pi,\pi)+\textbf{q}'$, with $|\mathbf q|,|\mathbf q'|\ll 1$. In this case, the energies
 can be approximated by Eq. (\ref{eq:quadratic dispersion})
and the vector $\textbf{h}(\textbf{k})$ by 
\begin{align}
\textbf{h}(\textbf{k}) & \approx(q_{x}q_{y},q_{y}q_{z},q_{x}q_{z})\equiv q^{2}\tilde{\textbf{h}}(\Omega),
\end{align}
where $\Omega$ is the solid angle in spherical coordinates. We can also approximate $A(\textbf{k}-\textbf{k}')\approx3$ and 
\bea
\mathscr{F}^{\eta}(\textbf{k},\textbf{k}')&\approx&1+\frac{\tilde{\textbf{h}}(\Omega)\cdot\tilde{\textbf{h}}(\Omega')}{|\tilde{\textbf{h}}(\Omega)||\tilde{\textbf{h}}(\Omega')|}.
\eea
Gathering all these approximations, we verify  that $(1/T_1)_\eta$ scales as  $T^{2}$ for $T\to0$, as could be anticipated  from the low-energy
density of states in Eq. (\ref{eq:density states}). A similar calculation assuming  momenta near the Fermi
lines leads to $(1/T_{1})_{\eta}\propto T^{3}$. While this result refers to the contribution from $\eta$ fermions, it also reflects the qualitative behavior of the total $1/T_{1}$ since 
 the dispersion relation of
the $\theta$ fermions is qualitatively similar.

\begin{figure}
\begin{centering}
\includegraphics[clip,width=0.9\columnwidth]{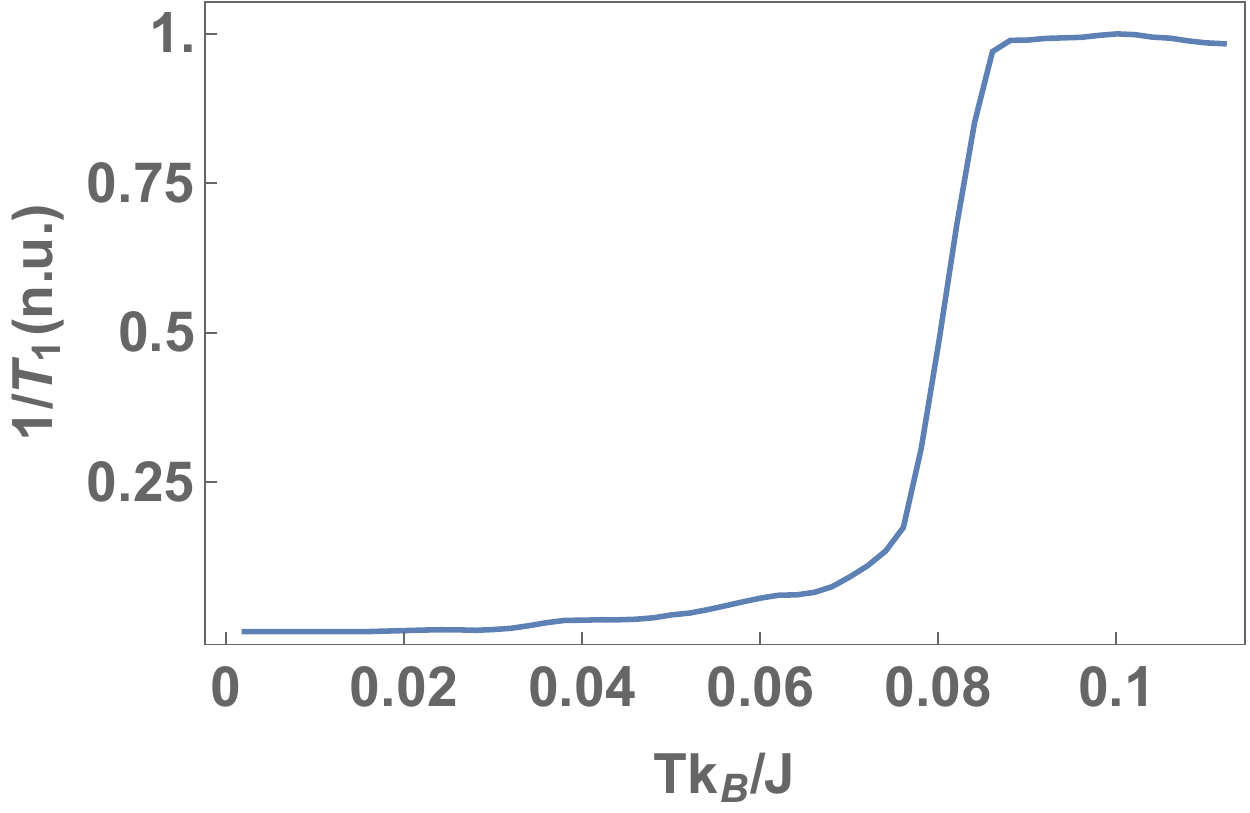} 
\par\end{centering}

\caption{\label{fig:Spin-lattice-relaxation}Spin lattice relaxation rate of
the chiral spin liquid state as a function of the temperature.}
\end{figure}

We have calculated  the total  spectral function $\chi_{+-}^{\prime\prime}(\textbf{q},\omega)$ numerically,
including the contribution from $\theta$ fermions and at arbitrary
temperatures, as explained in Appendix \ref{sec:Appendix}. The result
for the spin-lattice relaxation rate is shown in Fig. \ref{fig:Spin-lattice-relaxation}.
At low temperatures, the behavior is dominated by the $R$ point and
is described by  the power law $1/T_1\propto T^2$ discussed above. An abrupt increase
of $1/T_{1}$ is verified near the critical temperature $T_{c}$,
followed by a constant behavior at higher temperatures. This result
should be compared with figure 15(b) of Ref. \cite{Aharen2010}.
While the suppression of $1/T_1$ at low temperatures was interpreted as evidence for a gapped QSL, the experimental result is also qualitatively consistent with a pseudogap in the low-energy density of states. 
This  makes the chiral spin-orbital liquid state a valid alternative to explain the spin-lattice
relaxation rate of Ba$_{2}$YMoO$_{6}$.

\subsection{Inelastic Neutron Scattering \label{sec:INS}}

\begin{figure*}
\begin{centering}
\subfloat[\label{fig:DSF INS}]{\includegraphics[clip,width=0.5\columnwidth]{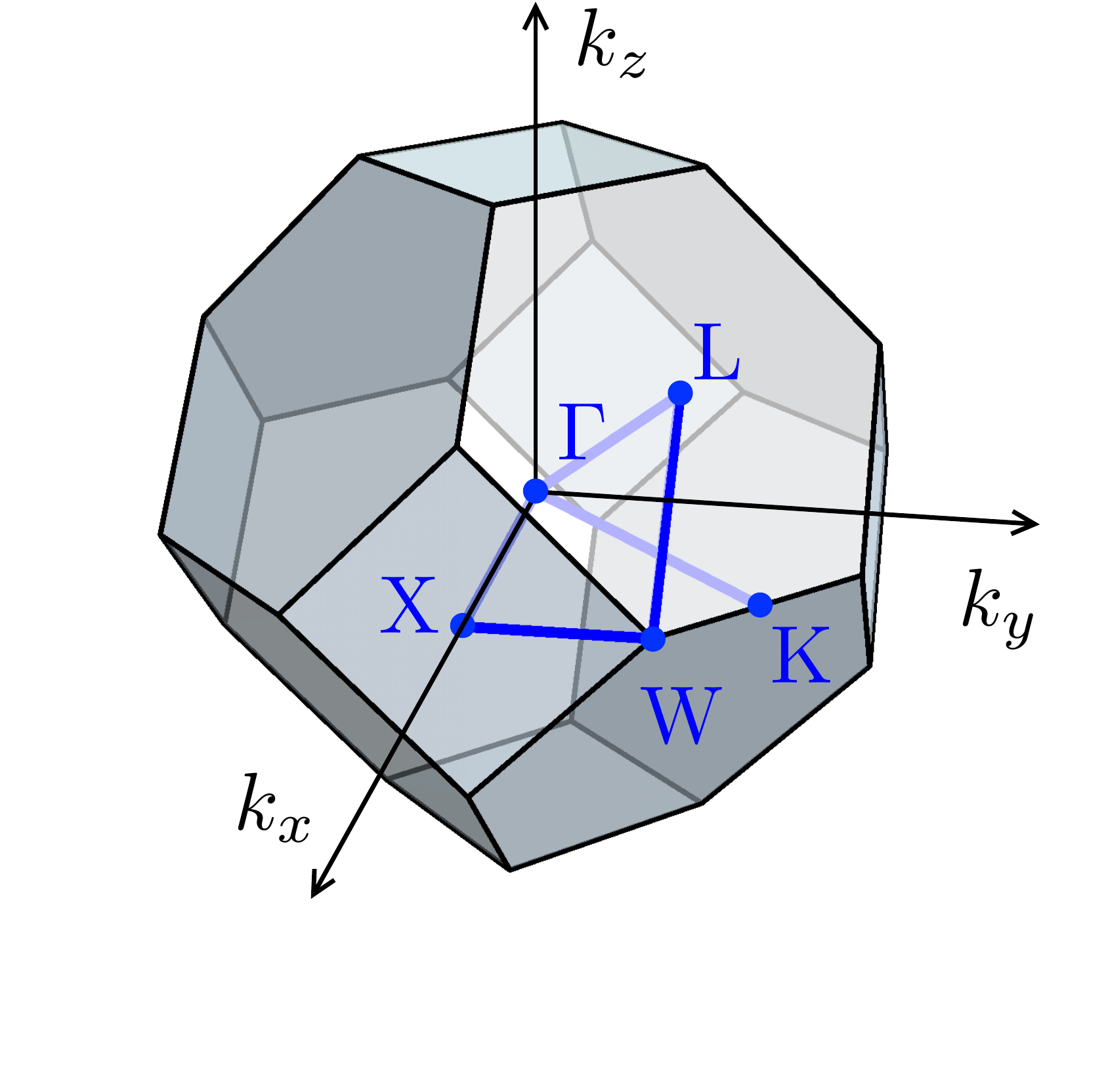}

}\subfloat[\label{fig:DSF INS fcc}]{\includegraphics[clip,width=0.6\columnwidth]{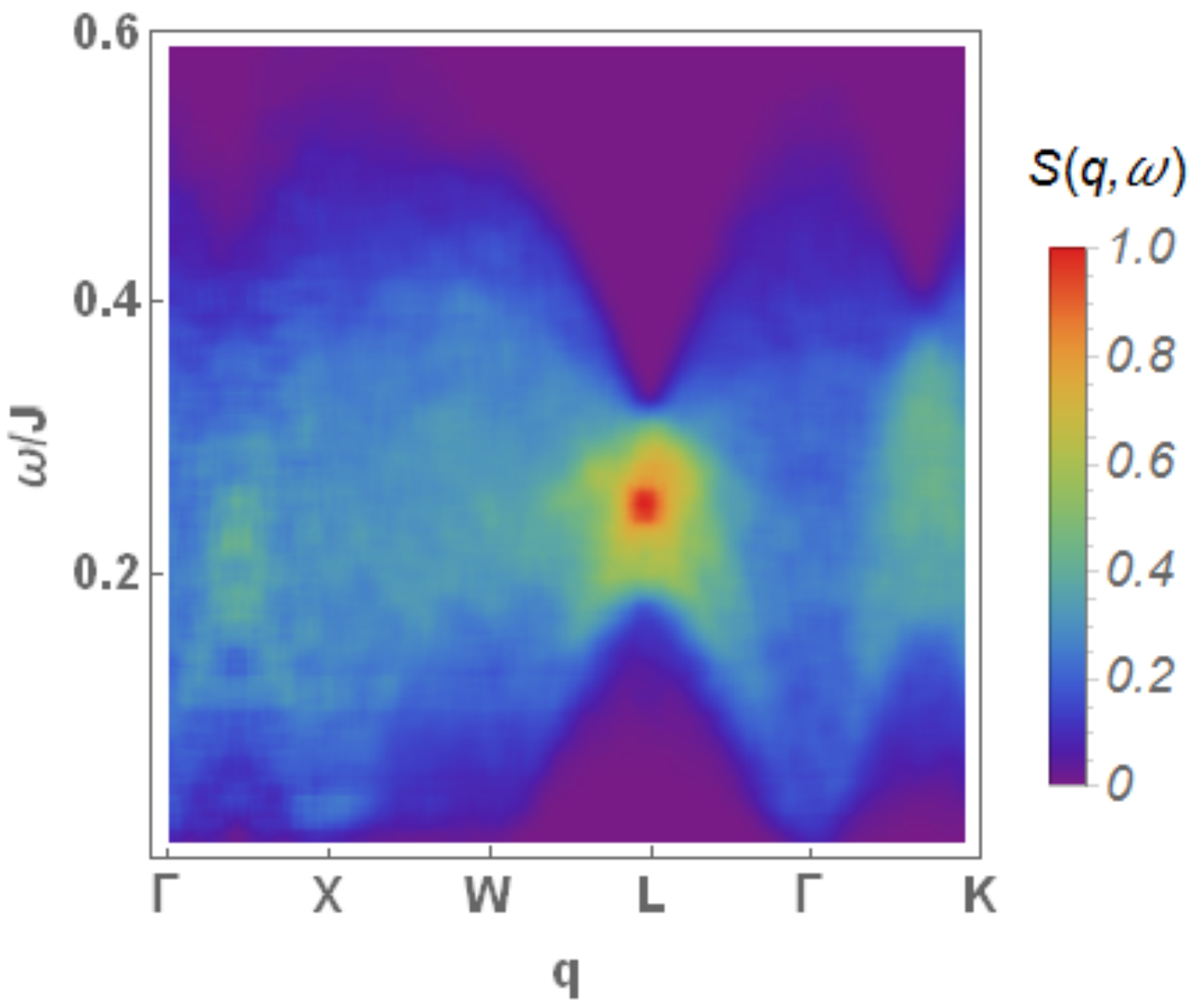}

}\hspace{1cm}\subfloat[\label{fig:DSF integrated}]{\includegraphics[clip,width=0.6\columnwidth]{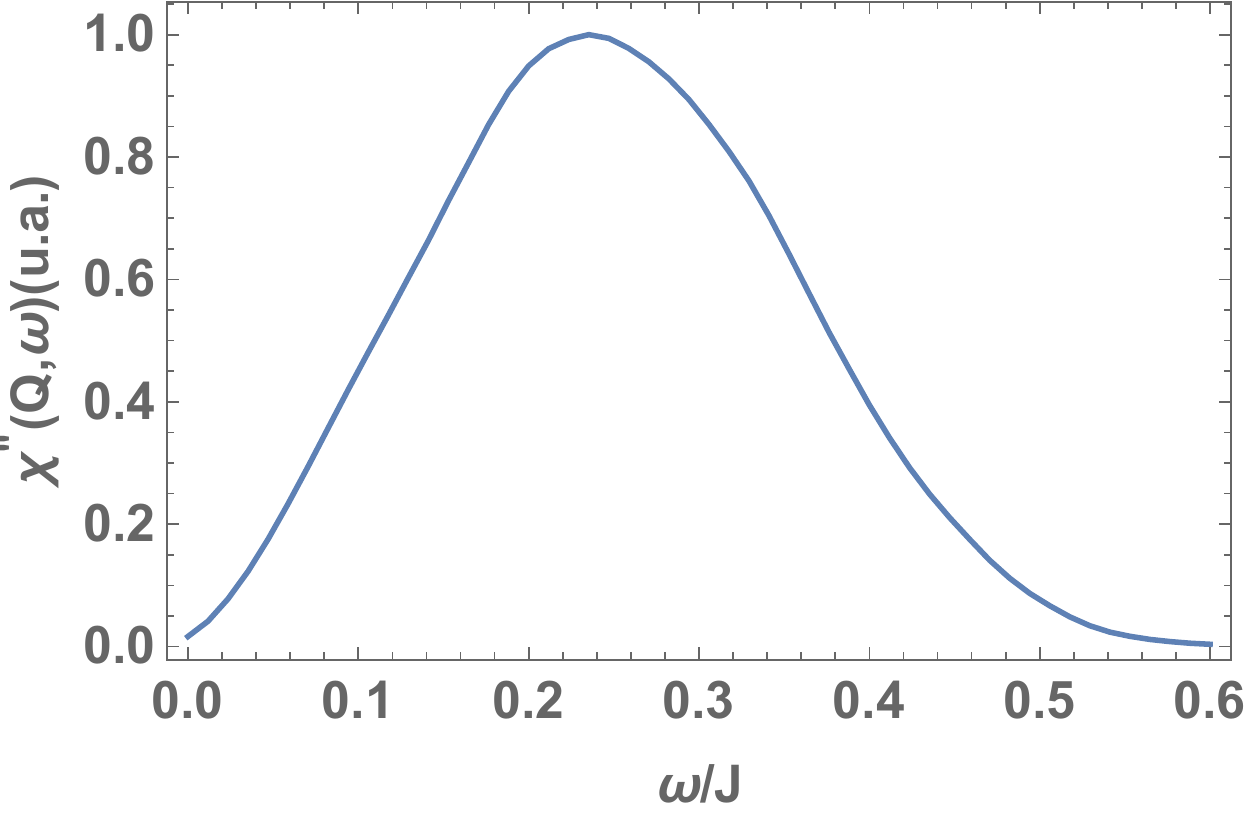}

}
\par\end{centering}

\caption{\label{fig:INS DynStrFac}(a) First Brillouin zone of the fcc lattice.
(b) Dynamical structure factor (in arbitrary units) probed by inelastic neutron scattering.
(c) Result after integration over 1.5\AA $^{-1}<Q<$1.8\AA $^{-1}$.}
\end{figure*}

Neutron scattering is the standard probe to study magnetic ordering
and excitations in condensed matter.  At zero temperature, the magnetic
scattering cross section  for polarized neutrons is proportional to one component of the dynamical structure factor 
\begin{align}
S^{ab}(\mathbf{q},\omega) & =\underset{j,n}{\sum}e^{-i\mathbf{q}\cdot\textbf{R}_{j}}\langle g\left|J_{j}^{a}\right|n\rangle\langle n\left|J_{i}^{b}\right|g\rangle\nonumber \\
 & \quad\times\delta(\omega-E_{n}+E_{g}),
\end{align}
where $|g\rangle$ is the ground state, $\mathbf q$ and $\omega>0$ are the momentum and   energy
transferred by the neutron, and $|n\rangle$ is an excited state of
the many-body Hamiltonian.

Here we will calculate the dynamical structure factor for the chiral spin-orbital liquid.  It follows from $PT$ and point group symmetries
that $S^{ab}(\mathbf{q},\omega)\propto\delta^{ab}$ and $S^{aa}(\mathbf{q},\omega)=S^{aa}(-\mathbf{q},\omega)$.
 Writing the operator $\textbf{J}$ in terms of $\textbf{s}$ and $\boldsymbol{\tau}$
and then  Majorana fermions, we obtain
\begin{equation}
S^{aa}(\mathbf{q},\omega)=\frac{4}{N}\underset{n,\mathbf k}{\sum}F_{n}^{a}(\mathbf{k},\mathbf{q})\delta(\omega-E_{n}+E_{g}).\label{eq:Saa}
\end{equation}
The form factor for $S^{zz}(\mathbf{q},\omega)$ is
\begin{align}
F_{n}^{z}(\mathbf{k},\mathbf{q}) =&\left|\underset{X}{\sum}\langle g|\eta_{\mathbf{q}-\mathbf{k}+\mathbf{G},X}^{2}\eta_{\mathbf{k},X}^{3}|n\rangle e^{i\mathbf{G}\cdot\boldsymbol{\delta}_{X}}\right|^{2}\nonumber \\
 & +4\left|\underset{X}{\sum}\langle g|\eta_{\mathbf{q}-\mathbf{k}+\mathbf{G},X}^{3}\bar{\theta}_{\mathbf{k},X}^{xy}|n\rangle e^{i\mathbf{G}\cdot\boldsymbol{\delta}_{X}}\right|^{2},\label{FFneutrons}
\end{align}
where $\mathbf{G}$ is a vector of the cubic reciprocal lattice chosen
such that $\mathbf{q}-\mathbf{k}+\mathbf{G}$ is contained in the
first Brillouin zone. The components $S^{xx}(\mathbf{q},\omega)$
and $S^{yy}(\mathbf{q},\omega)$ can be obtained from Eq. (\ref{FFneutrons})
by cyclic permutation of all indices $a=1,2,3=x,y,z$. Within
the mean-field theory, the excited states are restricted to two-particle
excitations. The form factors can be calculated using the matrix elements
of $U_{\mathbf{k}}$ and $V_{\mathbf{k}}$ defined in Eqs. (\ref{eq:defUk})
and (\ref{eq:defVk}).

Figure \ref{fig:DSF INS fcc} shows the dynamical structure factor
$S^{xx}(\mathbf{q},\omega)$ along the high-symmetry lines of the
Brillouin zone of the fcc lattice (see Fig. \ref{fig:DSF INS}). As
expected for QSLs, the spectral weight is distributed over a continuum
of fractionalized excitations. The maximum intensity is found at the
$L$ point, corresponding to momentum transfer $\mathbf{q}=(\pi,\pi,\pi)$.
The energy scale at the maximum is of the order of the bandwidth of
the Majorana fermions $\eta^{a}$ and $\theta^{1,3}$ shown in Fig.
\ref{fig:Dispersion fermions}. Using $J\approx72$ meV estimated
from the specific heat, we find that the peak in the dynamical structure
factor appears at $\omega\approx0.25J\approx18\text{ meV}$.

We now compare our theoretical results with the neutron scattering
experiments reported by Carlo \textit{et al.} \cite{Carlo2011} done
in polycrystalline samples. To make the comparison, we average the
dynamical structure factor over momenta with absolute value $Q$ in
the range 1.5\AA $^{-1}<Q<1.8$\AA $^{-1}$. This range includes the
point equivalent to $L$ called $L'=a^{-1}(3\pi,3\pi,-\pi)$, at which
$Q=1.63$\AA $^{-1}$ if we use the lattice spacing $a=8.389$\AA{}
\cite{deVries2010}. Our result in Fig. \ref{fig:DSF integrated}
shows a single broad maximum at $\omega\approx18$ meV. By contrast,
the experimental result shows a three-peak structure, with a pronounced
magnetic peak at $\omega\approx28$ meV and two smaller ones at 11
and 17 meV. As noted by the authors of Ref. \cite{Carlo2011} , the
energy scale of the broad peak is a factor of 2 larger than the one
inferred from the spin-lattice relaxation rate. 

Our result for the chiral spin-orbital liquid at the mean-field level
does not predict such a three-peak structure. Our model does contain
multiple energy scales associated with the nondegenerate Majorana fermion 
bands shown in Fig. \ref{fig:Dispersion fermions}, but the bandwidths
of the flavors $\eta^{a},\theta^{1,3}$, which appear in the form
factor Eq. (\ref{FFneutrons}), are rather close to each other. We
also recall that our calculations were done neglecting fluctuations
of the $Z_{2}$ gauge field and interactions between Majorana fermions.
The inclusion of these effects in a bosonic spin liquid on the kagome
lattice \cite{Punk2014} led to broadening and shift of the spectral
weight of $S(\textbf{q},\omega)$ when compared with the mean-field
theory. We expect a similar broadening in our case if gauge fluctuations
are taken into account.

\section{RIXS cross sections\label{sec:RIXS}}

RIXS is a photon-in photon-out spectroscopic technique that probes
excitations in solid state systems by measuring the energy, momentum
and polarization changes of the scattered photon \cite{Ament2011}.
It is a resonant technique because the x-ray is tuned
to coincide with the atomic transition between a core and a valence
level of a given atom. The resonance turns an otherwise negligible
second-order perturbation into the dominant contribution to the scattering
amplitude. Moreover, the transitions involved in the absorption
and emission processes are more complex than the ones generated by
the probes listed in Sec. \ref{sec:usual probes}, allowing for  the
experimental study of a manifold of elementary excitations. 

In this section, we evaluate and analyze the RIXS scattering operators
for cubic double perovskites and calculate the RIXS cross sections
for the chiral spin-orbital liquid. In subsection \ref{sub:RIXS-operators} we describe
the RIXS processes as well as  the approximations used in our calculation, and present a symmetry analysis of the scattering
operators using the method described in Refs. \cite{PhysRevB.81.125118,Haverkort2010,Savary2015}.
We stress that here the symmetry arguments
are applied to $j=3/2$ operators in the strong SOC limit and $O_{h}$
point-group symmetry. This is in contrast to Ref. \cite{Savary2015},
which focused on SU(2)-invariant spin-$1/2$ systems with negligible SOC. In Subsection \ref{sub:L-edge-RIXS}, we determine
the operators that appear specifically in the scattering amplitudes
for the $L$ edge. The RIXS cross sections for the chiral spin-orbital liquid are then calculated
and analyzed.

\subsection{Derivation and symmetry analysis of RIXS scattering operators\label{sub:RIXS-operators}}

Consider a general $N$ electron system, described by a many-body
Hamiltonian $H_{0}$. The total Hamiltonian describing the system
is $H=H_{0}+H^{\prime}$, where $H'$ describes the interaction between
electrons and photons 
\begin{equation}
H^{\prime}=\underset{i=1}{\overset{N}{\sum}}\left[\frac{e}{m}\textbf{A}(\textbf{r}_{i})\cdot\textbf{p}_{i}+\frac{e\hbar}{2m}\boldsymbol{\sigma}_{i}\cdot\nabla\times\textbf{A}(\textbf{r}_{i})\right].
\end{equation}
Concerning the electrons, $e$ is the charge, $m$ is the mass, $\textbf{r}_{i}$, and 
$\textbf{p}_{i}$ and $\boldsymbol{\sigma}_{i}$ are, respectively,
the position, momentum and spin of the $i$-th electron. The photon
is represented by the electromagnetic vector potential $\textbf{A}(\mathbf{r})$.
In second quantization, $\textbf{A}(\mathbf{r})$ is written as 
\begin{equation}
\textbf{A}(\textbf{r})=\underset{\textbf{k},\boldsymbol{\varepsilon}}{\sum}\frac{1}{\sqrt{2\mathcal{V}\epsilon_{0}\omega_{\textbf{k}}}}\left(\boldsymbol{\varepsilon}a_{\textbf{k},\boldsymbol{\varepsilon}}e^{i\textbf{k}\cdot\textbf{r}}+\boldsymbol{\varepsilon}^{\ast}a_{\textbf{k},\boldsymbol{\varepsilon}}^{\dagger}e^{-i\textbf{k}\cdot\textbf{r}}\right),
\end{equation}
where $\mathcal{V}$ is the volume, $\epsilon_{0}$ is the vacuum
permittivity, and $a_{\textbf{k},\boldsymbol{\varepsilon}}^{\dagger}$
is the creation operator for a photon with wave vector $\textbf{k}$,
frequency $\omega_{\mathbf{k}}$, and polarization vector  $\boldsymbol{\varepsilon}$. 

Our aim is to evaluate the x-ray scattering cross sections after
treating the photons as perturbations. Let the initial electron-photon
state be $|G\rangle$ and a set of final states be $\{|F\rangle\}$.
Using Fermi's golden rule to second order, we obtain the x-ray cross section
\begin{align}
I & \propto\underset{F}{\sum}\left|\langle F\left|H^{\prime}\right|G\rangle+\underset{\nu}{\sum}\frac{\langle F\left|H^{\prime}\right|\nu\rangle\langle\nu\left|H^{\prime}\right|G\rangle}{E_{G}-E_{\nu}+i\gamma_{\nu}}\right|^{2}\nonumber \\
 & \quad\,\,\times\delta(E_{F}-E_{G}),\label{eq:Golden rule}
\end{align}
in which $E_{\nu}$ and $1/\gamma_{\nu}$ are the energy and the lifetime
of the intermediate state $|\nu\rangle$, respectively. We assume
that the initial state corresponds to a direct product of a many-body
electronic ground state $|g\rangle$ and an incident photon state: $|G\rangle=|g\rangle\otimes|\mathbf{k},\omega_{\mathbf{k}},\boldsymbol{\varepsilon}\rangle$.
Similarly, the final state is a direct product of an excited electronic
state $|n\rangle$ with energy $E_n$ and an emitted photon labeled by $|F\rangle=|n\rangle\otimes|\mathbf{k}^{\prime},\omega_{\mathbf{k}^{\prime}},\boldsymbol{\varepsilon}^{\prime}\rangle$.
We also deal with the case where $\omega_{\textbf{k}}$ is tuned to
the energy difference between an atomic core level and a valence shell
state. The photon is totally absorbed and the $|\nu\rangle$ state
contains  an atomic core hole and an additional electron in the valence
or conduction band. If the photon energy $\hbar\omega_{\mathbf{k}}$
is tuned so that $|E_{G}-E_{\nu}|\ll\gamma_{\nu}$, the system is said
to be in resonance and the importance of second-order processes is
maximized.

Four standard approximations will be used to evaluate the second-order
terms in Eq. (\ref{eq:Golden rule}). First, we neglect the so-called
``magnetic'' contribution ($\propto\boldsymbol{\sigma}\cdot\nabla\times\textbf{A}$
) of $H^{\prime}$. Second, we   use the dipole approximation
for the scattering amplitude and take $e^{i\mathbf{k}\cdot\mathbf{r}_{i}}\approx e^{i\mathbf{k}\cdot\mathbf{R}_{i}}$,
where $\mathbf{R}_{i}$ represents the lattice point to which the $i$-th electron is bound. Third, we consider that the highly
unstable core hole in the $|\nu\rangle$ state decays before it 
can hop to a different ion. Finally, we consider only direct RIXS
processes, i.e., we neglect effects of the core-hole Coulomb potential
on the valence electrons. Within this fast collision approximation
\cite{Ament2011}, RIXS probes only single-site operators. The cross
section then depends only on $\mathbf{q}=\mathbf{k}-\mathbf{k}^{\prime}$
and $\omega=\omega_{\mathbf{k}}-\omega_{\mathbf{k}^{\prime}}$, which
are, respectively, the momentum and energy transferred to the sample.
Equation (\ref{eq:Golden rule}) can then be recast in the form  
\begin{equation}
I(\mathbf{q},\omega)\propto\underset{n}{\sum}|\langle n|\hat{\mathcal{O}}_{\mathbf{q}}|g\rangle|^{2}\delta(E_{g}-E_{n}+\hbar\omega),\label{eq:RIXS cross section}
\end{equation}
where $\hat{\mathcal{O}}_{\mathbf{q}}$ is the so-called scattering
operator in momentum space. The latter is obtained from  the Fourier transform
$\hat{\mathcal{O}}_{\mathbf{q}}=\underset{{i}}{\sum}e^{i\mathbf{q}\cdot\mathbf{R}_{i}}\hat{\mathcal{O}}_i$, where  
\begin{equation}
\hat{\mathcal{O}}_{i}=\underset{\nu}{\sum}\frac{1}{i\gamma_{\nu}}\mathcal{D}_{i}^{\dagger}(\boldsymbol{\varepsilon}^{\prime})|\nu\rangle\langle\nu|\mathcal{D}_{i}(\boldsymbol{\varepsilon}).\label{OofR}
\end{equation}
Here, the dipole operator 
\begin{equation}
\mathcal{D}_{i}(\boldsymbol{\varepsilon})=\boldsymbol{\varepsilon}\cdot\textbf{r}_{i}\label{eq:Dre}
\end{equation}
acts on the electronic states bound to  position  $\mathbf{R}_{i}$.

Equations (\ref{OofR}) and (\ref{eq:Dre}) show that the RIXS cross section
(\ref{eq:RIXS cross section}) depends on the initial
and final polarizations $\boldsymbol{\varepsilon}$
and $\boldsymbol{\varepsilon}^{\prime}$ and on the matrix elements of the electron position operator $\langle\nu|\mathbf{r}|g\rangle$
and $\langle n|\mathbf{r}|\nu\rangle$. The  photon polarizations can be   controlled in experiments (at
least in principle). However, the matrix elements   depend on   details of the intermediate states for a particular compound. The   general
claim one can make is that, provided the final states are low-energy
excitations, the scattering operators can be rewritten in terms of
charge, spin and orbital degrees of freedom of the valence electrons.
For magnetic insulators, RIXS  operators   correspond to a combination
of spin and orbital angular momentum. This feature makes
RIXS an attractive technique to investigate magnetic insulators with
strong SOC, in which spins and orbitals cannot be treated as separate
degrees of freedom.

Since $\hat{\mathcal{O}}_{i}$ is in general a   complicated
operator, it is desirable to  start our RIXS analysis by determining:
(i) which polarization vectors  $\boldsymbol{\varepsilon} $ and $\boldsymbol{\varepsilon}^{\prime}$ 
we should choose to acquire the signatures of a given state; and
(ii) which spin operators couple with these polarizations. The two
issues can be tackled at once by an elementary symmetry analysis of
Eq. (\ref{OofR}). As the absorption and emission processes occur
at the same ion, the operator $\hat{\mathcal{O}}_{i}$
should be invariant under operations of the point group symmetry of
the site $\mathbf{R}_{i}$. In general, one  starts by decomposing 
the scattering  operator into  irreducible representations of the point group, $\Gamma=\Gamma_{1}\oplus...\oplus\Gamma_{n}$.
A basis for these representations is then constructed in terms of
the polarization factors $\varepsilon^{\Gamma_{j}}$ and (pseudo)spins
$\mathcal{J}^{\Gamma_{j}}$, in the form \cite{PhysRevB.81.125118,Haverkort2010,Savary2015}
\begin{equation}
\hat{\mathcal{O}}_{i}=\underset{\Gamma_{j}=1}{\overset{n}{\sum}}\underset{l_{j}=1}{\overset{\text{mul}(\Gamma_{j})}{\sum}}\kappa_{\Gamma_{j},l_{j}}\varepsilon^{\Gamma_{j},l_{j}}\cdot\mathcal{J}^{\Gamma_{j},l_{j}},\label{eq:Tr}
\end{equation}
where $\text{mul}(\Gamma_{j})$ is the multiplicity of the irreducible
representation $\Gamma_{j}$, the dot represents a symmetric contraction
of all indices, and $\kappa_{\Gamma_{j},l_{j}}$ are material specific
coefficients.

The bases of the irreducible representations of the octahedral group
in terms of multipoles of $j=3/2$ moments are known
\cite{Santini2009,Shiina1998}  and are reproduced in Table \ref{tab:multipoles},
together with their representation in terms of $\textbf{s}$ and $\boldsymbol{\tau}$
pseudospins. It is also easy to verify that the following
polarization factors form the bases $\varepsilon^{\Gamma_{j},l_{j}}$:
\begin{subequations}\label{eq:pol fac}\begin{align}
P_{a} & =\frac{i}{2}\underset{bc}{\sum}\epsilon_{abc}\varepsilon_{b}^{\prime\ast}\varepsilon_{c},  \\
T_{a} & =\frac{1}{2}\underset{b\neq c}{\sum}(1-\delta_{ab})(1-\delta_{ac})\varepsilon_{b}^{\prime\ast}\varepsilon_{c},  \\
Q_{2} & =\varepsilon_{x}^{\prime\ast}\varepsilon_{x}-\varepsilon_{y}^{\prime\ast}\varepsilon_{y},  \\
Q_{3} & =\frac{1}{\sqrt{3}}(\varepsilon_{x}^{\prime\ast}\varepsilon_{x}+\varepsilon_{y}^{\prime\ast}\varepsilon_{y}-2\varepsilon_{z}^{\prime\ast}\varepsilon_{z}),  \\
U & =\boldsymbol{\varepsilon}^{\prime\ast}\cdot\boldsymbol{\varepsilon},
\end{align}
\end{subequations}
Here, the vector 
$\textbf{P}$ corresponds to the $\Gamma_{4}$ representation, $\textbf{T}$ to  
$\Gamma_{5}$, $ Q_2$ and $Q_3$ to   $\Gamma_{3}$, and $U$ to the
scalar representation. Combining the operators in the same irreducible representation in Table \ref{tab:multipoles} with the polarization factors in Eqs.  
(\ref{eq:pol fac}) according to Eq. (\ref{eq:Tr}), we find the general form of
all transition operators {[}except for the scalar representation,
which couples with the Casimir operator $\textbf{J}^{2}=j(j+1)=\text{const.}${]}. 
We  then see that 
RIXS can in principle  directly probe  pseudospin and pseudo-orbital  excitations. 

\subsection{L-edge RIXS cross section of Mo$^{5+}$\label{sub:L-edge-RIXS}}

\begin{figure}
\includegraphics[width=0.9\columnwidth]{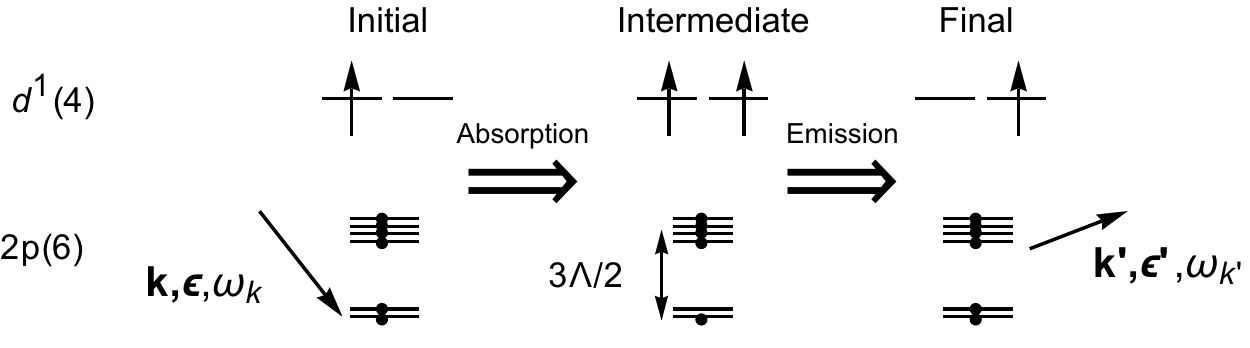}

\caption{\label{fig:RIXS scheme}Schematic diagram of a RIXS experiment at
the $L_{2}$ edge, featuring specifically the possibility of pseudo-orbital
flip. The absorbed photon creates a core $2p$ hole, which is subject
to  strong spin-orbit coupling. This highly unstable state  decays before a $d$ electron can tunnel to or from the ion, generating a spin-orbital  excitation and an emitted photon.}
\end{figure}

We now focus on the $L$-edge RIXS operators for $4d^{1}$ and $5d^{1}$
orbital systems retaining cubic symmetry, whose mechanism is illustrated
in Fig. \ref{fig:RIXS scheme}. At the $L$ edge \cite{Ament2011},
2$p$ core electrons are excited to the $B$ and $C$ states of Eq.
(\ref{eq:ABC}). To describe the core-hole states, we first note that
they are similar to the $t_{2g}$ valence states, since they   result from the combination of 
spin-$1/2$ states with orbital angular momentum $L=1$ in the presence of SOC.
The core-hole Hamiltonian is 
\begin{equation}
H_{\text{core}}=\Lambda\textbf{L}\cdot\textbf{S},\label{eq:Hcore}
\end{equation}
where $\Lambda>0$ is the SOC constant for the $2p$ states. Like
in the $d^{1}$ valence electron, there is a lifting of the six-fold
degeneracy into a doublet and a quadruplet. However, now the doublet
has lower energy (see Fig. \ref{fig:RIXS scheme}). We refer to the 
excited hole   in the $j=1/2$ ($j=3/2$) multiplet as the resonant  $L_{2}$ ($L_{3}$) edge.

\begin{figure*}
\subfloat[\label{fig:CSa-1}]{\includegraphics[clip,width=0.5\columnwidth]{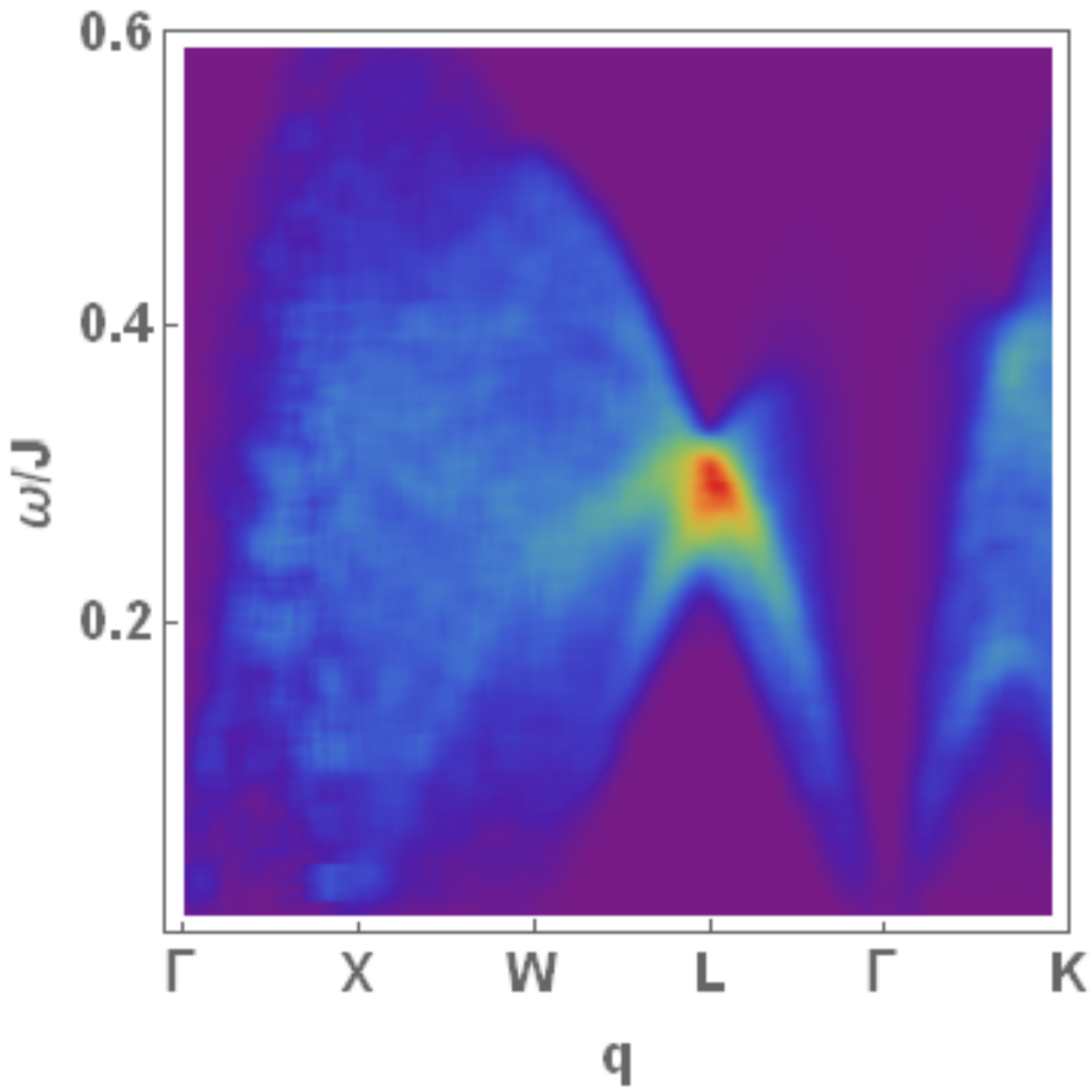}

} \subfloat[\label{fig:CSb-1}]{\includegraphics[clip,width=0.5\columnwidth]{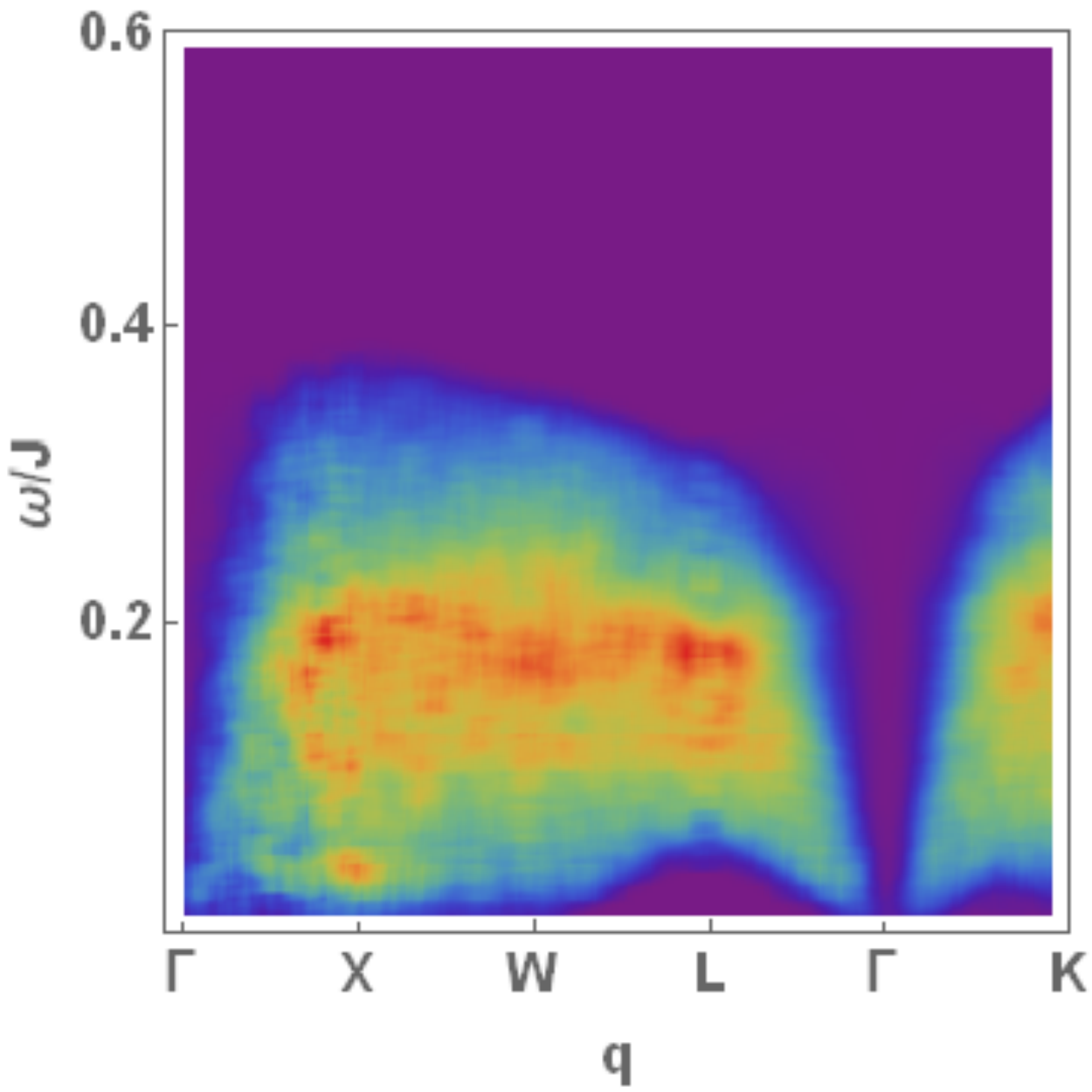}

} \subfloat[\label{fig:CSc-1}]{\includegraphics[clip,width=0.5\columnwidth]{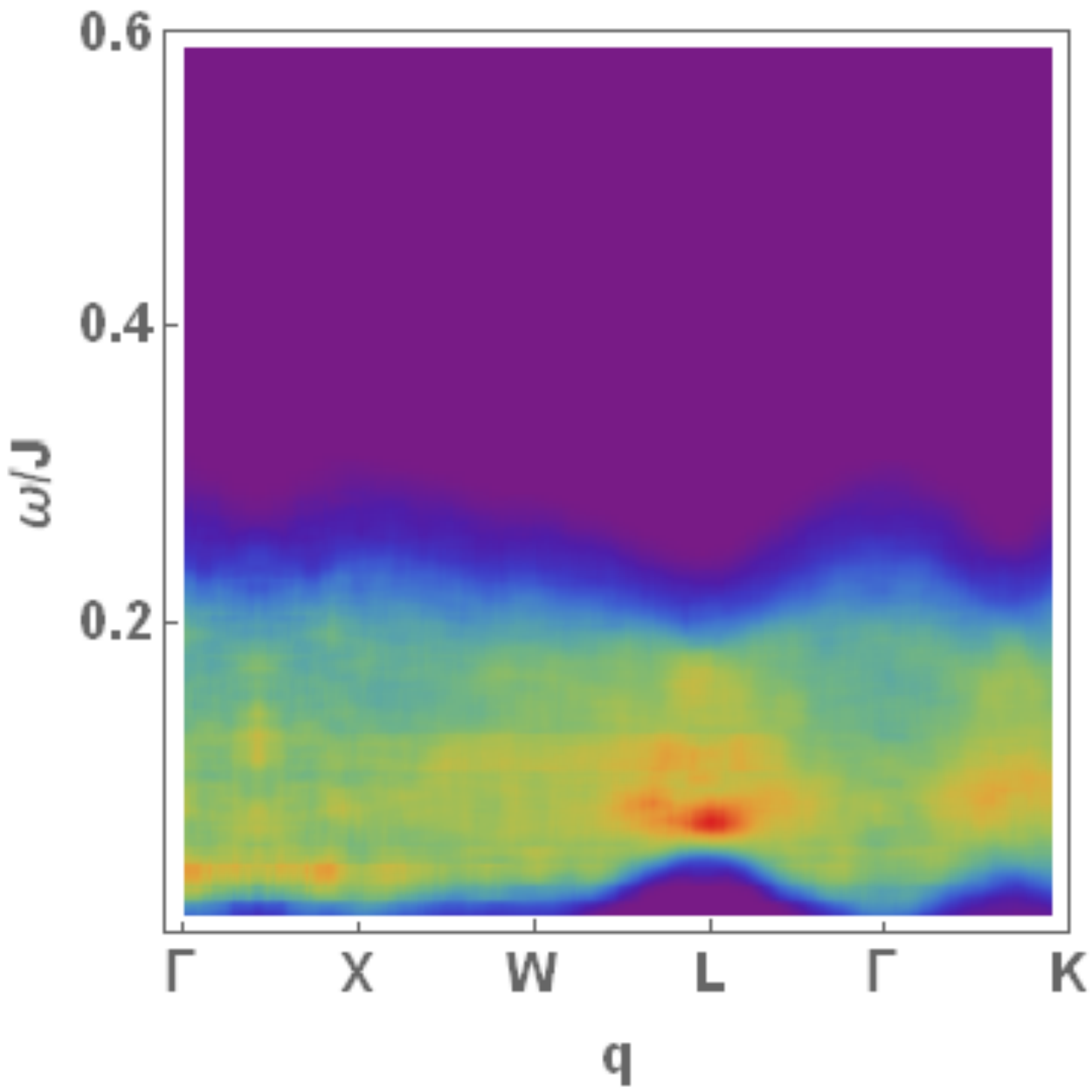}

} \subfloat[\label{fig:CSd-1}]{\includegraphics[clip,width=0.59\columnwidth]{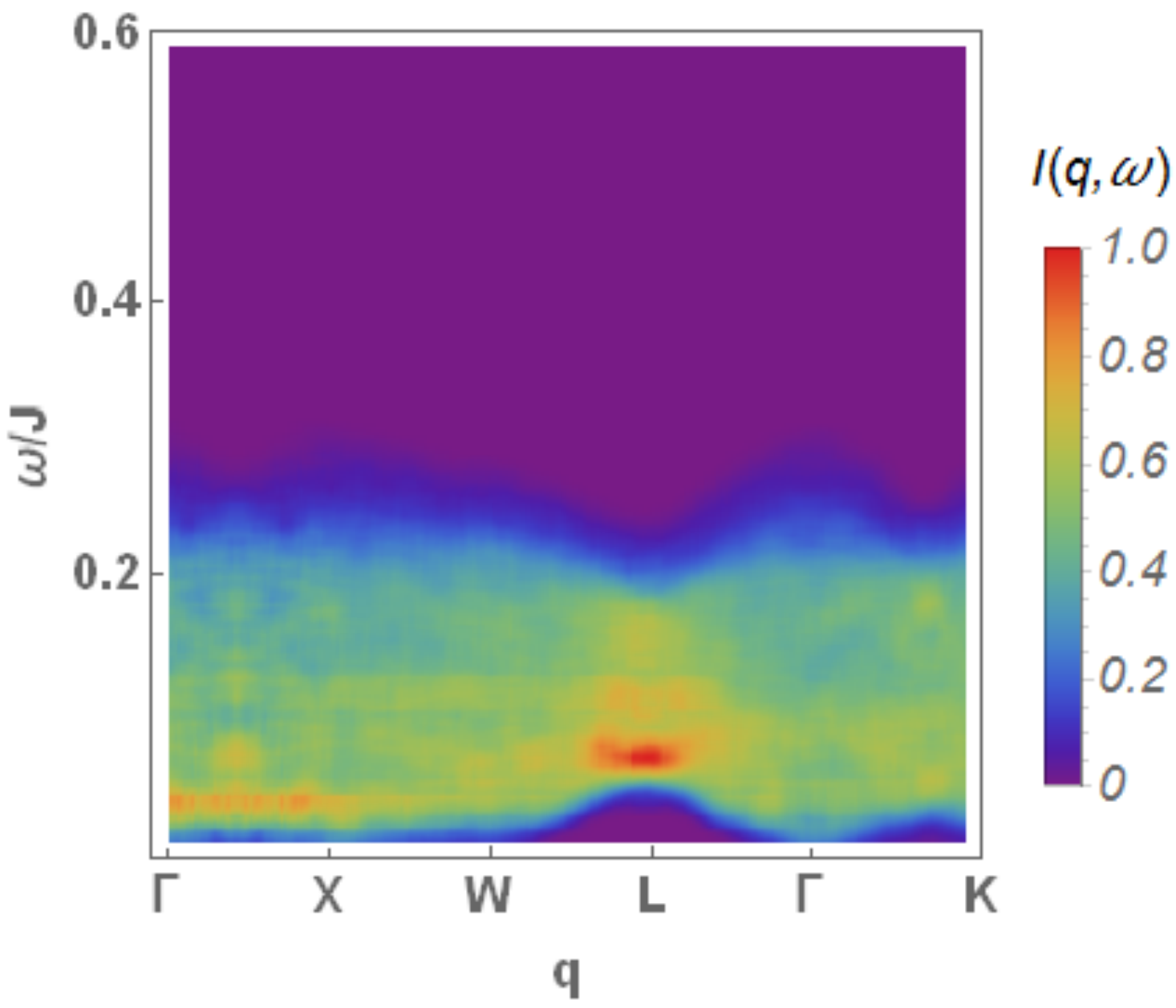}

}

\caption{\label{fig:RIXS cross section} RIXS cross section (in arbitrary units)
probing (a) $\textbf{s}$, (b) $\textbf{s}\tau_{y}$, (c) $\tau_{x}$
and (d) $\tau_{z}$ operators along the high symmetry directions of
the Brillouin zone of the fcc lattice.}
\end{figure*}

Most of the dipole matrix elements of Eq. (\ref{eq:Dre}) vanish by
symmetry. The remaining terms are written in second quantization as
\cite{Kim2017} 
\begin{align}
\mathcal{D}_{i}(\boldsymbol{\varepsilon})\propto  &d_{xy,\sigma}^{\dagger}(\varepsilon_{x}p_{y,\sigma}+\varepsilon_{y}p_{x,\sigma}) +d_{yz,\sigma}^{\dagger}(\varepsilon_{y}p_{z,\sigma}+\varepsilon_{z}p_{y,\sigma})\nonumber \\
 & +d_{zx,\sigma}^{\dagger}(\varepsilon_{z}p_{x,\sigma}+\varepsilon_{x}p_{z,\sigma}),
\end{align}
in which we have dropped a multiplicative factor $\langle4d_{yz}|y|2p_{z}\rangle$.
We can simplify Eq. (\ref{OofR}) by writing $\gamma_{\nu}\approx\gamma_{\mu}=\text{const.}$,
with $\mu=2,3$, for all intermediate states in the $L_{\mu}$
edge. Here, $\gamma_{\mu}$ is the   average decay rate 
of the intermediate core-hole states. This approximation, together
with the ones  discussed in Section \ref{sub:RIXS-operators}, leads to
\begin{align}
\hat{\mathcal{O}}_{i}^{L_\mu}= & \frac{1}{i\gamma_{\mu}}\mathcal{D}_{i}^{\dagger}(\boldsymbol{\varepsilon}^{\prime})\mathcal{P}_{\mu}\mathcal{D}_{i}(\boldsymbol{\varepsilon}),\label{eq:Ozeta}
\end{align}
where $\mathcal{P}_{\mu}$ is the projection operator of the intermediate
states in  the $L_{\mu}$ edge.  

We can derive expressions for the scattering operator in terms of pseudospins $\mathbf s$ and $\mathbf \tau$ by taking the projection in the $j=3/2$ subspace and using the single-occupancy  constraint $\sum_{\sigma}(B_{\sigma}^{\dagger}B_{\sigma}+C_{\sigma}^{\dagger}C_{\sigma})=1$. In Appendix \ref{sec:Table-scattering-operators}, we provide a general
expression for Eq. (\ref{eq:Ozeta}) including the effects of a tetragonal
distortion that lifts the degeneracy between $B$ and $C$ states. Here, we restrict the discussion to the cubic
limit. For the $L_{2}$ edge, we find
\begin{align}
\hat{\mathcal{O}}_i^{L_2}  \propto&\frac{1}{\sqrt3}\left[Q_{2}\tau_i^{x}+Q_{3}\tau_i^{z}-4\mathbf{T}\cdot(\mathbf{s}\tau_i^{y})-\frac{2}{\sqrt3}\mathbf{P}\cdot\mathbf{K}_i\right]\nonumber\\
&+\text{const.},\label{O2R}
\end{align}
where
\begin{equation}
\mathbf{K}_i=\left(s_i^{x}(1-4\tau_i^{yz}),s_i^{y}(1-4\tau_i^{xz}),s_i^{z}(1-4\tau_i^{xy})\right).
\end{equation}
On the other hand, the scattering operator for the $L_{3}$ edge involves only the pseudospin $\mathbf s$: 
\begin{equation}
\hat{\mathcal{O}}^{L_3}_i\propto\frac{4}{3}\mathbf{P}\cdot\mathbf{s}_i+\text{const.}.\label{O3R}
\end{equation}

After calculating the scattering operators $\hat{\mathcal{O}}_i^{L_\mu}$,
the RIXS cross section in Eq. (\ref{eq:RIXS cross section}) can be
calculated like the INS dynamical structure factor discussed in Section
\ref{sec:INS}. The results for some representative operators are
shown in Fig. \ref{fig:RIXS cross section}. Once again, we find that
the spectral weight is distributed over a broad continuum. A common
feature for all   these results     is a maximum of intensity
for transferred momentum at the $L$ point, $\textbf{q}=(\pi,\pi,\pi)$.

Interestingly, the $\theta^{2}$ fermion is excited in the cross section
of the $L_{2}$ edge through the operators $\tau^{x}$, $\tau^{z}$
and $\textbf{s}\tau^{y}$, in sharp contrast with the dynamical structure
factor for INS. Due to the reduced bandwidth of the $\theta^{2}$
fermions, the spectrum probed by RIXS (with the proper polarization)
  displays a narrower energy range when compared to the one measured
by INS. This feature is readily verified when comparing Figs. \ref{fig:CSb-1},
\ref{fig:CSc-1} and \ref{fig:CSd-1} with Fig. \ref{fig:DSF INS fcc}.

Let us turn to the $L_{3}$ edge, which detects pseudospin excitations
directly. We can simplify the result by  choosing $\boldsymbol{\varepsilon}$
and $\boldsymbol{\varepsilon}^{\prime}$ such that $P_{x}=P_{y}=0$,
but $P_{z}\neq0$. The cross section in this case is given by 
\begin{equation}
I(\textbf{q},\omega)\propto\underset{n}{\sum}\left|\left\langle n\left|s_{\textbf{q}}^{z}\right|g\right\rangle \right|^{2}\delta(E_{g}-E_{n}+\hbar\omega),\label{eq:I rixs L3}
\end{equation}
where $s_{\textbf{q}}^{z}$ is the Fourier transform of $s_{j}^{z}$.
At the special point $\textbf{q}=0$, the form factor involves the conserved quantity $\textbf{s}_{\textbf{q}=0}=\textbf{s}_{\text{tot}}$
[see Eq. (\ref{eq:total pseudospin})], which commutes with the spin Hamiltonian. Since the ground state
is a singlet of the pseudospin  SU(2) symmetry, we have
$\textbf{s}_{\textbf{q}=0}|g\rangle=0$. 
 Thus, it follows from Eq. (\ref{eq:I rixs L3}) that 
\begin{equation}
I(\mathbf{q}=0,\omega)=0\quad\text{(for }\hat{\mathcal{O}}_{\mathbf{q}}=s_{\mathbf{q}}^{z}),\label{eq:limit q=00003D00003D0}
\end{equation}
for any transferred energy $\omega$. This feature is clearly seen
in Fig. \ref{fig:CSa-1}, and should be contrasted with the dynamical structure
factor $S(\mathbf{q}=0,\omega)\neq0$ for INS in Fig. \ref{fig:DSF INS fcc}.
This result is explicitly confirmed by the computation of the form
factor in Eq. (\ref{eq:I rixs L3}). At the mean field level, the
excited state $\left|n\right\rangle $ is a two-particle excitation,
in which the particles are characterized by well-defined momenta $\textbf{k}$
and $\textbf{k}^{\prime}$. We can write $\left|n\right\rangle = \left|n(\mathbf{k},\mathbf{k}^{\prime})\right\rangle$, in which the vector $\mathbf{k}^{\prime}$ can
take the values $\pm\mathbf{k}\pm\mathbf{q}$ according to the type of two-particle
excitation under consideration. As shown in Appendix \ref{sec:Appendix}
(see Eq. (\ref{eq:RIXS form appendix})), the form factor in this
case is 
\begin{equation}
\left|\left\langle n(\mathbf{k},\mathbf{k}^{\prime})\left|s_{\mathbf{q}}^{z}\right|g\right\rangle \right|^{2}=1-\frac{\mathbf{h}(\mathbf{k})\cdot\mathbf{h}(\mathbf{k}^{\prime})}{|\mathbf{h}(\mathbf{k})||\mathbf{h}(\mathbf{k}^{\prime})|}.\label{eq:RIXS form factor}
\end{equation}
This form factor clearly vanishes for $\mathbf{q}=0$. Therefore,
this RIXS cross section could be used to detect the hidden SU(2) symmetry
of the spin-orbital model for double perovskites.

We note that the   dynamic structure factor for the operator $\textbf{s}\tau^{y}$ calculated at mean-field level also vanishes at  $\mathbf  q=0$ (see Fig. \ref{fig:CSb-1}).  The reason is that the Majorana representation
 $s^{a}\tau^{y}=-i\eta^{a}\theta^{2}/4$, involves only   $\theta^{2}$ and   $ \eta$ fermions, whose mean-field Hamiltonian is 
 diagonalized by the same unitary transformation $U_{\mathbf{k}}$
given by Eq. (\ref{eq:defUk}). As a result, the form factor associated
with $\textbf{s}\tau^{y}$ is also given by Eq. (\ref{eq:RIXS form factor}). However, since $\sum_{j}\textbf{s}_{j}\tau_{j}^{y}$ does not commute
with the Hamiltonian, the vanishing  of the spectral weight at $\mathbf q=0$ in this case is an artifact of the mean-field
approximation.

We make here a final remark on the usefulness of RIXS to probe our
QSL. Our discussion was restricted to one-site operators, but this
technique can, in principle, probe operators involving two or three
sites. Extending the symmetry arguments presented here,
we predict that the chiral operator $\textbf{s}_{i}\cdot(\textbf{s}_{j}\times\textbf{s}_{k})$
could be probed and would couple with the polarization factor $\boldsymbol{\varepsilon}^{\prime\ast}\cdot\boldsymbol{\varepsilon}$
(see Table I of Ref. \cite{Savary2015}). Therefore, RIXS could in
principle detect the $P$ and $T$ symmetry breaking of the chiral
spin-orbital liquid in the elastic limit. Once again, we emphasize
that our results were obtained within the parton mean-field theory.
The role of $Z_{2}$ gauge fluctuations in the RIXS response deserves
a separate and detailed study.

\section{Conclusions\label{sec:Conclusions}}

This paper presented a theoretical study of thermodynamic and spectroscopic
properties of a $j=3/2$ Majorana chiral spin-orbital liquid. In the
process, we fully developed a pseudospin representation of the $d^{1}$
orbital physics in Mott insulators with strong SOC. These results
can guide the theoretical modeling, as well as the design and interpretation
of experiments in compounds with similar local physics. Interestingly,
the thermodynamic properties of the chiral spin-orbital liquid agree qualitatively
with the available experimental results for the material Ba$_2$YMoO$_6$  \cite{deVries2010,deVries2013,Carlo2011,Aharen2010}.
In particular, we find a sharp drop in the spin-lattice relaxation
rate at low temperatures, even though the chiral spin-orbital liquid
is a gapless phase. On the other hand, the inelastic neutron scattering
cross section measured in Ref. \cite{Carlo2011} was not reproduced,
since we found a single broad peak instead of the three-peak structure observed 
in polycrystalline samples. Adding effects beyond mean-field theory
may explain this difference and will be left for future work.

As the main result of this paper, we showed that RIXS can selectively
probe pseudospin and pseudo-orbital operators, and thus provide a
direct way to detect quadrupolar and octupolar orders and excitations.
Our results   give some guidance
to interpret RIXS spectra in 4$d^{1}$ and 5$d^{1}$ based compounds.
In particular, we showed that the hidden SU(2) symmetry of the double perovskite model without  Hund's coupling can be demonstrated by probing pseudospin $\textbf{s}$ excitations and observing  the suppression of the spectral weight for momentum transfer at the $\Gamma$ point.

Finally, we note that the analysis of RIXS scattering operators studied here is also useful for other ordered double perovskites \cite{Chen2010,Romhanyi2017,Svoboda2017}. For magnetically ordered systems, the excitation spectrum can be fitted using  a microscopic model
[such as Eq. (\ref{eq:total Hamiltonian}) in Appendix \ref{sec:Hund-Coupling-Terms}] and representing spin-orbital excitations in terms of magnons within a spin-wave theory \cite{Li1998,Joshi1999}.  For instance, the onset  of quadrupolar order in some osmium-based compounds observed   in a recent study \cite{Lu2017} can be investigated in more
detail using  RIXS. Two recent RIXS measurements of spin waves, one in a
compound preserving cubic structure \cite{Betto2017} and another in a $j=3/2$ compound \cite{Jeong2017}, indicate 
that the theory developed in this paper can be tested in the near future.

We thank E. Andrade, F. A. Garcia, G. Jackeli, and E. Miranda for
helpful discussions. This work was supported by Brazilian agencies
FAPESP (W.M.H.N.) and CNPq (R.G.P.).

\appendix

\section{Orbital physics with distortion\label{sec:Orbital-Physics}}

Throughout the main text, we kept our discussion of the orbital physics restricted to 
case of   the cubic symmetry. In this appendix, we discuss the effects of tetragonal
distortions on the $A,B,C$ states. The Hamiltonian (\ref{eq:Hion})
is redefined by 
\begin{equation}
H_{\text{ion}}=-\lambda\textbf{\textbf{l}}\cdot\textbf{S}+\delta(l^{z})^{2},
\end{equation}
where $\delta$ is the energy scale associated with the distortion.
Notice that the time-reversal symmetry of $H_{\text{ion}}$ is preserved,
which means that the eigenstates can still be organized into three Kramers
pairs. In analogy with Eq. (\ref{eq:ABC}), we define

\begin{align}
A_{\sigma} & =2\sigma\left(\sin\varphi\,d_{0,-\sigma}-\cos\varphi\,d_{-2\sigma,\sigma}\right),\nonumber \\
B_{\sigma} & =\cos\varphi\,d_{0,-\sigma}+\sin\varphi\,d_{-2\sigma,\sigma},\nonumber \\
C_{\sigma} & =d_{2\sigma,\sigma},\label{eq:ABC distortion}
\end{align}
in which the angle $\varphi$ is defined by \begin{equation}
\tan(2\varphi)=\frac{2\sqrt{2}\lambda}{\lambda+2\delta}.\label{phiangle}\end{equation}
The
corresponding  energies are given by 
\begin{align}
\epsilon_{A(B)} & =\frac{1}{2}\left[\frac{\lambda}{2}+\delta\pm\sqrt{\left(\frac{\lambda}{2}+\delta\right)^{2}+2\lambda^{2}}\right],\nonumber \\
\epsilon_{C} & =-\frac{\lambda}{2}+\delta,\label{eq:energies distortion}
\end{align}
showing how distortion lifts the degeneracy of the cubic limit. Notice
that the $\textbf{s} $ and $\boldsymbol{\tau}$ operators can still be
used to describe the physics of the quadruplet formed by the orbitals
$B$ and $C$. 

\section{Effective Hamiltonian for nonzero  Hund's coupling  \label{sec:Hund-Coupling-Terms}}

In the limit of strong SOC, we can project the Hamiltonian in Eq. (\ref{eq:Heff without SOC})
in the $j=3/2$ manifold as written in Eq. (\ref{eq:projection}).  Here we present the more general  effective Hamiltonian for $\eta\neq0$.  We introduce the pseudo-orbital-dependent operators: 
\begin{align}
\tilde{\mathcal{S}}_{ij}^{\alpha\beta} & =\left(\frac{1}{2}-\tau_{i}^{\alpha\beta}\right)\left(\frac{1}{2}-\tau_{j}^{\alpha\beta}\right),\\
\tilde{\mathcal{Q}}_{ij}^{\alpha\beta} & =\sqrt{3}\left(\frac{1}{2}-\tau_{i}^{\alpha\beta}\right)\bar{\tau}_{j}^{\alpha\beta}+\left(i\leftrightarrow j\right),\\
\tilde{\mathcal{R}}_{ij}^{\alpha\beta} & =\left(\frac{1}{2}-\tau_{i}^{\alpha\beta}\right)(1+\tau_{j}^{\alpha\beta})+\left(i\leftrightarrow j\right),
\end{align}
in which $\bar{\tau}^{\alpha\beta}=\frac{1}{\sqrt{3}}(\tau^{\beta\gamma}-\tau^{\gamma\alpha})$.
The Hamiltonian   is given by 
\begin{align}
H_{\text{eff}} =& \frac{4}{9}J\underset{\langle ij\rangle_{\gamma}}{\sum}\left(\textbf{s}_{i}\cdot\textbf{s}_{j}+\frac{1}{4}\right)\tilde{\mathcal{S}}_{ij}^{\alpha\beta}+\frac{4}{9}V\underset{\langle ij\rangle_{\gamma}}{\sum}\tilde{\mathcal{S}}_{ij}^{\alpha\beta},\nonumber\\
&-\frac{4}{9}J^{\prime}\underset{\langle ij\rangle_{\gamma}}{\sum}\left[(s_{i}^{\alpha}s_{j}^{\alpha}-s_{i}^{\beta}s_{j}^{\beta})\tilde{\mathcal{Q}}_{ij}^{\alpha\beta}  -s_{i}^{\gamma}s_{j}^{\gamma}\tilde{\mathcal{R}}_{ij}^{\alpha\beta}\right]\nonumber \\
 &+\frac{2}{3}J^{\prime}\underset{\langle ij\rangle_{\gamma}}{\sum}\tilde{\mathcal{S}}_{ij}^{\alpha\beta}.\label{eq:total Hamiltonian}
\end{align}
The coupling constants $J$, $J'$ and $V$ are defined by Eqs. (\ref{eq:Jcoupling}), (\ref{eq:Jpcoupling}) and (\ref{eq:Vcoupling}).

\section{Free energy near the critical point\label{sec:Landau's free energy}}

In this appendix, we find an approximate expression for Eq. (\ref{eq:Gen formula free energy})
near the finite-temperature critical point where the order parameters  of the parton mean-field theory  vanish. Expanding (\ref{eq:Gen formula free energy})
up to the fourth-order in $\beta\epsilon_{\textbf{k}\lambda}$, we
find 
\begin{align}
\Phi\equiv&\frac{\beta F}{N}\nonumber\\
=&-3\ln2+\frac{\mc K}{2}\left(u^{2}+u\bar{w}+\frac{vw}{3}\right)\nonumber \\
 & -3\left(\frac{\mc K}{36}\right)^{2}\left(21u^{2}+v^{2}+w^{2}+12u\bar{w}+3\bar{w}^{2}\right)\nonumber \\
 & +\frac{3}{8}\left(\frac{\mc K}{36}\right)^{4}\left[19(3u+v)^{4}+84(2u+\bar{w})^{4}\right.\nonumber \\
 & \left.-24uv\left(11(3u+v)^{2}-39uv\right)+28w^{4}\right],\label{eq:F}
\end{align}
where $\mc K=\beta J$. We reorganize $\Phi$   in the form \[
\Phi\equiv-3\ln2+ \Phi_2(u,v,w,\bar{w})+\Phi_{4}(u,v,w,\bar{w}),
\]
where $\Phi_2$ contains the terms that are quadratic in the order parameters and $\Phi_4$ contains the quartic terms. 
The quadratic term can be written in   matrix
form $\Phi_2=\textbf{t}^{T}M\textbf{t}$, where $\textbf{t}^{T}=(u,v,w,\bar{w})$.
Diagonalizing $M$, we find the set of eigenvalues $a_n$, $n=1,\dots,4$, given by 
\begin{align}
a_{1,2} & =-\frac{\mc K(36\pm \mc K)}{432},\nonumber \\
a_{3,4} & =\frac{\mc K(36-4\mc K\pm\sqrt{2592-360\mc K+13\mc K^{2}})}{144},
\end{align}
The eigenvalues  $a_{2}$ and $a_{4}$ vanish, respectively, at the temperatures
$k_BT_{p}=J/36$ and $k_BT_{c}=J/12$. The critical temperature  where the numerically calculated specific heat in Fig. \ref{fig:Magnetic-specific-heat} drops to zero corresponds to the higher value $T=T_{c}$.

\begin{table*}
\caption{\label{tab:Coefficients} Coefficients of the scattering operators in Eq.
\ref{eq:RIXS scattering distortion} as a function of the angle parameter
$\varphi$. The columns with the cubic limit values are obtained by
taking $\varphi=\arcsin(1/\sqrt{3})$.}

\begin{tabular}{|c|c|c||c|c|}
\hline
&$L_{3}$ edge &Cubic &$L_{2}$ edge& Cubic\\
\hline 
$a_{\mu,U}$ & $\frac{1}{9}\left(\cos^{2}\varphi-\sqrt{2}\sin2\varphi+3\sin^{2}\varphi-1\right)$ & 0 & $\frac{1}{9}\left(\cos^{2}\varphi+\sqrt{2}\sin2\varphi-2\right)$ & 0\tabularnewline
\hline 
$a_{\mu,Q_{2}}$ & $\frac{\sqrt{3}}{18}\left(\cos^{2}\varphi-\sqrt{2}\sin2\varphi+2-6\sin^{2}\varphi\right)$ & 0 & $\frac{\sqrt{3}}{18}\left(\cos^{2}\varphi+\sqrt{2}\sin2\varphi+4\right)$ & $\frac{\sqrt{3}}{3}$\tabularnewline
\hline 
$a_{\mu,Q_{3}}$ & $2\sqrt{2}\left(\cos\varphi-\sqrt{2}\sin\varphi\right)$ & 0 & $2\left(\sqrt{2}\cos\varphi+\sin\varphi\right)$ & $\frac{\sqrt{3}}{3}$\tabularnewline
\hline 
$a_{\mu,T_{x}}$ & $-\frac{2\sqrt{2}}{3}\left(\cos\varphi-\sqrt{2}\sin\varphi\right)$ & 0 & $-\frac{4}{3}\left(\sqrt{2}\cos\varphi+\sin\varphi\right)$ & $-\frac{4\sqrt{3}}{3}$\tabularnewline
\hline 
$a_{\mu,T_{y}}$ & $-\frac{2\sqrt{2}}{3}\left(\cos\varphi-\sqrt{2}\sin\varphi\right)$ & 0 & $-\frac{4}{3}\left(\sqrt{2}\cos\varphi+\sin\varphi\right)$ & $-\frac{4\sqrt{3}}{3}$\tabularnewline
\hline 
$a_{\mu,T_{z}}$ & $\frac{4\sqrt{2}}{3}\left(\cos\varphi-\sqrt{2}\sin\varphi\right)$ & 0 & $-\frac{4}{3}\left(\sqrt{2}\cos\varphi+\sin\varphi\right)$ & $-\frac{4\sqrt{3}}{3}$\tabularnewline
\hline 
$a_{\mu,P_{x}}$ & $\frac{1}{3}\left(2+\frac{3\sqrt{2}}{2}\sin2\varphi\right)$ & $\frac{4}{3}$ & $-\frac{2}{3}$ & $-\frac{2}{3}$\tabularnewline
\hline 
$a_{\mu,P_{y}}$ & $\frac{1}{3}\left(2+\frac{3\sqrt{2}}{2}\sin2\varphi\right)$ & $\frac{4}{3}$ & $-\frac{2}{3}$ & $-\frac{2}{3}$\tabularnewline
\hline 
$a_{\mu,P_{z}}$ & $\frac{1}{3}\left(4\cos^{2}\varphi+\sqrt{2}\sin2\varphi\right)$ & $\frac{4}{3}$ & $-\frac{1}{3}\left(\cos^{2}\varphi+\sqrt{2}\sin2\varphi\right)$ & $-\frac{2}{3}$\tabularnewline
\hline 
$b_{\mu,P_{x}}$ & $\frac{2\sqrt{2}}{3}\left(\cos\varphi-\sqrt{2}\sin\varphi\right)$ & 0 & $\frac{4}{3}(\mbox{\ensuremath{\sqrt{2}}}\cos\varphi+\sin\varphi)$ & $\frac{4\sqrt{3}}{3}$\tabularnewline
\hline 
$b_{\mu,P_{y}}$ & $-\frac{2\sqrt{2}}{3}\left(\cos\varphi-\sqrt{2}\sin\varphi\right)$ & 0 & $-\frac{4}{3}(\mbox{\ensuremath{\sqrt{2}}}\cos\varphi+\sin\varphi)$ & $-\frac{4\sqrt{3}}{3}$\tabularnewline
\hline 
$b_{\mu,P_{z}}$ & 0 & 0 & $0$ & 0\tabularnewline
\hline 
$c_{\mu,P_{x}}$ & $\frac{1}{3}\left(4-3\sqrt{2}\sin2\varphi\right)$ & 0 & $-\frac{4}{3}$ & $-\frac{4}{3}$\tabularnewline
\hline 
$c_{\mu,P_{y}}$ & $\frac{1}{3}\left(4-3\sqrt{2}\sin2\varphi\right)$ & 0 & $-\frac{4}{3}$ & $-\frac{4}{3}$\tabularnewline
\hline 
$c_{\mu,P_{z}}$ & $-\frac{2}{3}\left(\sqrt{2}\sin2\varphi-4\sin^{2}\varphi\right)$ & 0 & $\frac{1}{3}\left(5+\cos2\varphi+2\sqrt{2}\sin2\varphi\right)$ & $\frac{8}{3}$\tabularnewline
\hline 
\end{tabular}
\end{table*}

\section{Computation of correlation functions\label{sec:Appendix}}

In this appendix, we outline the calculation of   finite-temperature spectral functions such as the one in  Eq. (\ref{spectralchi}). 

We start by considering the correlation function \begin{equation}
\chi_{lm}(\tau)=\langle  T_\tau\hat O _l(\tau)\hat O_m(0)\rangle, 
\end{equation}
where $\hat O_l$ is a local operator acting on the $j=3/2$ subspace associated with site $l$,   $\hat O_l(\tau)=e^{H_{\text{eff}}\tau}\hat O_le^{-H_{\text{eff}}\tau}$ is the operator evolved in imaginary time, $T_\tau$ denotes time ordering, and  $\langle\cdot \rangle =\text{Tr}(\rho\,\cdot)$ denotes the thermal average with density matrix $\rho=e^{-\beta H_{\text{eff}}}/Z$.

Quite generally, the local operator  $\hat O_l$ can be written as a combination of Majorana fermion bilinears, $\zeta^a_l\zeta^b_l$, with $\zeta^{a}\in\{\eta^a, \theta^a\}$.  Let us illustrate the procedure by taking \be
\hat O_l=-i\eta_l^1\eta_l^2=2 s^z_l.
\ee  
Within the mean-field approximation, the correlation function can be written as\bea
\chi_{lm}(\tau)&=&\mathscr G_{ml}^{12}(-\tau)\mathscr G_{lm}^{21}(\tau)-\mathscr G_{ml}^{11}(-\tau)\mathscr G_{lm}^{22}(\tau),\label{correlation}\eea
where \be
\mathscr G_{lm}^{ab}(\tau)=-\langle T_\tau \eta_l^a(\tau)\eta_m^b(0)\rangle.
\ee
is the  noninteracting fermion Green's function. 
If $\mathbf R_l$ belongs to the $X$ sublattice, $X=1,\dots,4$, and $\mathbf R_m$ to the $Y$ sublattice, we can write  for $0<\tau<\beta$ [using momentum conservation and Eq. (\ref{eq:Uk eigen})]\bea
\mathscr G_{lm}^{ab}(\tau)&=&-\frac{8}N\sum_{\mathbf k\in \frac12\text{BZ}}\sum_\lambda\left[(U_{\mathbf k})_{X\lambda }(U^\dagger_{\mathbf k})_{\lambda Y}\langle \eta^a_{\mathbf k\lambda}\eta^b_{-\mathbf k\lambda}\rangle\right.\nonumber\\
&&\times e^{i\mathbf k\cdot (\mathbf R_l-\mathbf R_m)}e^{-\epsilon_{\mathbf k\lambda}^{(\eta)}\tau}+(U_{\mathbf k})_{Y\lambda }(U^\dagger_{\mathbf k})_{\lambda X} \nonumber\\
&&\left.\times \langle \eta^a_{-\mathbf k\lambda}\eta^b_{\mathbf k\lambda}\rangle e^{-i\mathbf k\cdot (\mathbf R_l-\mathbf R_m)}e^{\epsilon_{\mathbf k\lambda}^{(\eta)}\tau}\right].\label{Gfermion}
\eea
The thermal average yields \bea
\langle \eta^a_{\mathbf k\lambda}\eta^b_{-\mathbf k\lambda}\rangle&=&\delta^{ab}n_F(-\epsilon^{(\eta)}_{\mathbf k\lambda}).
\eea
Taking the Fourier transform of Eq. (\ref{Gfermion}), we obtain\bea
\mathscr G(\mathbf k,\omega_n)&=&\int_0^\beta d\tau \,e^{i\omega_n\tau}\frac1N\sum_{l,m}e^{-i\mathbf k\cdot (\mathbf R_l-\mathbf R_m) }\mathscr G_{lm}^{11}(\tau)\nonumber\\
&=&\frac12\sum_{X,Y}\frac{(U_{\mathbf k})_{X\lambda }(U^\dagger_{\mathbf k})_{\lambda Y}}{i\omega_n-\epsilon^{(\eta)}_{\mathbf k\lambda}},
\eea
where    $\omega_n=(2n+1)\pi /\beta$, $n\in\mathbb Z$, are fermionic Matsubara frequencies. 

Similarly, we obtain the Fourier transform of the correlation in Eq. (\ref{correlation}):\bea
\chi(\mathbf q,\omega_m)&=&\frac1N\sum_{\mathbf k\in\frac12 \text{BZ}}\sum_{\lambda_1,\lambda_2}\nonumber\\
&&\times\left\{\frac{\mc F^{(1)}_{\lambda_1\lambda_2}(\mathbf k,\mathbf q) [n_F(\epsilon^{(\eta)}_{\mathbf k-\mathbf q,\lambda_1})-n_F(\epsilon^{(\eta)}_{\mathbf k,\lambda_2})]}{i\omega_m- \epsilon^{(\eta)}_{\mathbf k\lambda_2}+ \epsilon^{(\eta)}_{\mathbf k-\mathbf q,\lambda_1}}\right.\nonumber\\
&&+\frac{\mc F^{(2)}_{\lambda_1\lambda_2}(\mathbf k,\mathbf q) [n_F(-\epsilon^{(\eta)}_{-\mathbf k+\mathbf q,\lambda_1})-n_F(\epsilon^{(\eta)}_{\mathbf k,\lambda_2})]}{i\omega_m- \epsilon^{(\eta)}_{\mathbf k\lambda_2}- \epsilon^{(\eta)}_{-\mathbf k+\mathbf q,\lambda_1}}\nonumber\\
&&+\frac{\mc F^{(3)}_{\lambda_1\lambda_2}(\mathbf k,\mathbf q) [n_F(\epsilon^{(\eta)}_{\mathbf k,\lambda_2})-n_F(\epsilon^{(\eta)}_{\mathbf k+\mathbf q,\lambda_1})]}{i\omega_m- \epsilon^{(\eta)}_{\mathbf k+\mathbf q,\lambda_1}+\epsilon^{(\eta)}_{\mathbf k\lambda_2}}\nonumber\\
&&\left.+\frac{\mc F^{(4)}_{\lambda_1\lambda_2}(\mathbf k,\mathbf q) [n_F(\epsilon^{(\eta)}_{\mathbf k,\lambda_2})-n_F(-\epsilon^{(\eta)}_{-\mathbf k-\mathbf q,\lambda_1})]}{i\omega_m-  \epsilon^{(\eta)}_{-\mathbf k-\mathbf q,\lambda_1}+\epsilon^{(\eta)}_{\mathbf k\lambda_2}}\right\},\nonumber\\
&&
\eea
where $\omega_m=2\pi m/\beta$, $m\in\mathbb Z$,  are bosonic Matsubara frequencies. 
The form factors are given by \bea
\mathcal{F}_{\lambda_{1}\lambda_{2}}^{(1)}(\textbf{k},\textbf{q}) & =&\left|\underset{X}{\sum}e^{i\textbf{G}\cdot\boldsymbol{\delta}_{X}}\left(U_{\textbf{k}-\textbf{q}+\mathbf G}^{\dagger}\right)_{\lambda_{1}X}\left(U_{\textbf{k}}\right)_{X\lambda_{2}}\right|^{2},  \nonumber\\
\mathcal{F}_{\lambda_{1}\lambda_{2}}^{(2)} (\textbf{k},\textbf{q})& =&\left|\underset{X}{\sum}e^{i\textbf{G}\cdot\boldsymbol{\delta}_{X}}\left(U_{-\textbf{k}+\textbf{q}+\mathbf G}^{\phantom\dagger}\right)_{X\lambda_{1}}\left(U_{\textbf{k}}^{\phantom\dagger}\right)_{X\lambda_{2}}\right|^{2},\nonumber \\
\mathcal{F}_{\lambda_{1}\lambda_{2}}^{(3)}(\textbf{k},\textbf{q}) & =&\left|\underset{X}{\sum}e^{i\textbf{G}\cdot\boldsymbol{\delta}_{X}}\left(U_{\textbf{k}}^{\dagger}\right)_{\lambda_{2}X}\left(U_{\textbf{k}+\textbf{q}+\mathbf G}^{\phantom\dagger}\right)_{X\lambda_{1}}\right|^{2},\nonumber \\
\mathcal{F}_{\lambda_{1}\lambda_{2}}^{(4)} (\textbf{k},\textbf{q})& =&\left|\underset{X}{\sum}e^{i\textbf{G}\cdot\boldsymbol{\delta}_{X}}\left(U_{-\textbf{k}-\textbf{q}+\mathbf G}^{\dagger}\right)_{\lambda_{1}X}\left(U_{\textbf{k}}^{\dagger}\right)_{\lambda_{2}X}\right|^{2},\nonumber\\
&&
\eea
where $\mathbf G=2\pi(n_x,n_y,n_z)$ with $n_a\in \mathbb Z$ are reciprocal lattice vectors chosen such that the momenta $\pm \mathbf k\pm \mathbf q+\mathbf G$ in each form factor  lies in $\frac12$BZ. 

After an analytical continuation $i\omega_m\to \omega+i0^+$,  we can take the imaginary part of the retarded correlation function $\chi^{\prime\prime}(\mathbf q,\omega)$ in a standard way. In the regime $\beta \omega\ll 1$, we can approximate the factors of Fermi-Dirac distributions using $n_{F}(\epsilon+\omega)-n_{F}(\epsilon)   \approx\omega dn_{F}/d\epsilon$. We then obtain the expression for the contribution from the $\eta$ fermions to the spin-lattice relaxation rate 

\begin{align}
\left(\frac{1}{T_{1}}\right)_{\eta}\propto & \frac{\pi}{4N}\sum_{\lambda_{1},\lambda_{2}}\,\underset{\textbf{k}\in\frac{1}{2}\text{BZ}}{\sum}\,\underset{\textbf{q}\in\text{BZ}}{\sum}\frac{|A(\textbf{q})|^{2}}{\cosh^{2}\left(\beta\epsilon_{\textbf{k}\lambda_{2}}^{(\eta)}/2\right)}\times\nonumber \\
 & \times\left[\mathcal{F}_{\lambda_{1}\lambda_{2}}^{(1)}(\textbf{k},\textbf{q})\delta(\epsilon_{\textbf{k}-\textbf{q},\lambda_{1}}^{(\eta)}-\epsilon_{\textbf{k}\lambda_{2}}^{(\eta)})\right.\nonumber \\
 & +\mathcal{F}_{\lambda_{1}\lambda_{2}}^{(2)}(\textbf{k},\textbf{q})\delta(\epsilon_{-\textbf{k}+\textbf{q},\lambda_{1}}^{(\eta)}+\epsilon_{\textbf{k}\lambda_{2}}^{(\eta)})\nonumber \\
 & +\mathcal{F}_{\lambda_{1}\lambda_{2}}^{(3)}(\textbf{k},\textbf{q})\delta(\epsilon_{\textbf{k}+\textbf{q},\lambda_{1}}^{(\eta)}-\epsilon_{\textbf{k}\lambda_{2}}^{(\eta)})\nonumber \\
 & \left.+\mathcal{F}_{\lambda_{1}\lambda_{2}}^{(4)}(\textbf{k},\textbf{q})\delta(\epsilon_{-\textbf{k}-\textbf{q},\lambda_{1}}^{(\eta)}+\epsilon_{\textbf{k},\lambda_{2}}^{(\eta)})\right]\label{eq:1T1 appendix}.
\end{align}
Eq. (\ref{eq:1T1 appendix}) can be further simplified since $|\epsilon_{\textbf{k}\lambda_{i}}^{(\eta)}|=|\epsilon_{\textbf{k}\lambda_{j}}^{(\eta)}|$
for $\lambda_{i},\lambda_{j}=1,...,4$ (see Eq. (\ref{eq:eta dispersion})).
For $\mathcal{F}_{\lambda_{1}\lambda_{2}}^{(1)}(\textbf{k},\textbf{q})$,
the sum over eigenstates yields 

\begin{align}
\sum_{\lambda_{1},\lambda_{2}}\mathcal{F}_{\lambda_{1}\lambda_{2}}^{(1)}(\textbf{k},\textbf{q})\delta(\epsilon_{\textbf{k}-\textbf{q},\lambda_{1}}^{(\eta)}-\epsilon_{\textbf{k}\lambda_{2}}^{(\eta)}) & =2\mathscr{F}^{\eta}(\mathbf{k},\mathbf{k}-\mathbf{q})\nonumber \\
 &\quad \times\delta(|\epsilon_{\textbf{k}-\textbf{q}}^{(\eta)}|-|\epsilon_{\textbf{k}}^{(\eta)}|),\label{eq:Fappendix}
\end{align}
where 
\begin{equation}
\mathscr{F}^{\eta}(\mathbf{k},\mathbf{k}-\mathbf{q})=1+\frac{\mathbf{h}(\mathbf{k})\cdot\mathbf{h}(\mathbf{k}-\mathbf{q})}{|\mathbf{h}(\mathbf{k})||\mathbf{h}(\mathbf{k}-\mathbf{q})|}.\label{eq:NMR form factor 2}
\end{equation}
The expressions obtained for other sums differ from (\ref{eq:Fappendix})
only by the combination of vectors $\mathbf{k}$ and $\mathbf{q}$.
Eq. (\ref{eq:NMR form factor 2}) is the form factor $\mathscr{F}^{\eta}$ stated in Eq.
(\ref{eq:NMR form factor}). Notice that the reciprocal lattice vector
$\mathbf{G}$ does not appear in this final expression.

From $\chi(\textbf{q},\omega)$ we can also recover the RIXS dynamical
structure factor of $s^{z}$. Taking the zero temperature limit, we
find 
\begin{align}
\underset{T\rightarrow0^{+}}{\lim}\chi^{\prime\prime}(\textbf{q},\omega)\propto & \frac{\pi}{4N}\sum_{\lambda_{1},\lambda_{2}}\,\underset{\textbf{k}\in\frac{1}{2}\text{BZ}}{\sum}\,\underset{\textbf{q}\in\text{BZ}}{\sum}\underset{i}{\sum}\nonumber \\
 & \times\mathcal{F}_{\lambda_{1}\lambda_{2}}^{(i)}(\textbf{k},\textbf{q})\delta_{\lambda_{1}\lambda_{2}}^{(i)}(\omega,\textbf{k},\textbf{q}),
\end{align}
in which

\begin{align}
\delta_{\lambda_{1}\lambda_{2}}^{(1)}(\omega,\textbf{k},\textbf{q}) & =\Theta(-\epsilon_{\lambda_{1}})\Theta(\epsilon_{\lambda_{2}})\delta(\omega-(\epsilon_{\textbf{k}\lambda_{2}}^{(\eta)}-\epsilon_{\textbf{k}-\textbf{q},\lambda_{1}}^{(\eta)})),\nonumber \\
\delta_{\lambda_{1}\lambda_{2}}^{(2)}(\omega,\textbf{k},\textbf{q}) & =\Theta(\epsilon_{\lambda_{1}})\Theta(\epsilon_{\lambda_{2}})\delta(\omega-(\epsilon_{-\textbf{k}+\textbf{q},\lambda_{1}}^{(\eta)}+\epsilon_{\textbf{k}\lambda_{2}}^{(\eta)})),\nonumber \\
\delta_{\lambda_{1}\lambda_{2}}^{(3)}(\omega,\textbf{k},\textbf{q}) & =\Theta(\epsilon_{\lambda_{1}})\Theta(-\epsilon_{\lambda_{2}})\delta(\omega-(\epsilon_{\textbf{k}+\textbf{q},\lambda_{1}}^{(\eta)}-\epsilon_{\textbf{k}\lambda_{2}}^{(\eta)})),\nonumber \\
\delta_{\lambda_{1}\lambda_{2}}^{(4)}(\omega,\textbf{k},\textbf{q}) & =\Theta(-\epsilon_{\lambda_{1}})\Theta(-\epsilon_{\lambda_{2}})\delta(\omega+\epsilon_{-\textbf{k}-\textbf{q},\lambda_{1}}^{(\eta)}+\epsilon_{\textbf{k},\lambda_{2}}^{(\eta)}).
\end{align}
Once again, summing over the eigenstates, we find for $\omega=0$
\begin{align}
\sum_{\lambda_{1},\lambda_{2}}\mathcal{F}_{\lambda_{1}\lambda_{2}}^{(1)}(\textbf{k},\textbf{q})\delta_{\lambda_{1}\lambda_{2}}^{(1)}(0,\textbf{k},\textbf{q}) & =\left(1-\frac{\textbf{h}(\textbf{k})\cdot\textbf{h}(\textbf{k}-\textbf{q})}{|\textbf{h}(\textbf{k})||\textbf{h}(\textbf{k}-\textbf{q})|}\right)\nonumber \\
 &\quad \times\delta(|\epsilon_{\textbf{k}-\textbf{q}}^{(\eta)}|-|\epsilon_{\textbf{k}}^{(\eta)}|),\label{eq:RIXS form appendix}
\end{align}
with similar expressions for other summations. The expression in brackets
is just the form factor written in Eq. (\ref{eq:RIXS form factor}).
As stated in the main text, it is clear that this form factor will
vanish when $\mathbf{q}=0$.

The procedure outlined in this appendix can be generalized for the
$\theta$ fermions as well. In particular, if $\hat{O}_{l}=-i\eta_{l}^{a}\theta_{l}^{2}$,
the corresponding form factor of a RIXS experiment will be exactly
the one given in Eq. (\ref{eq:RIXS form appendix}). For operators
involving the fermions $\theta^{1}$ and $\theta^{3}$, it is not
possible to find exact expressions to the form factors, since
there is no closed form to the matrix $V_{\mathbf{k}}$ (see Eq. \ref{eq:defVk}).
The response functions must then be computed numerically.

\section{RIXS scattering operators \label{sec:Table-scattering-operators}}

In this appendix, we present the RIXS scattering operators discussed
in Section \ref{sec:RIXS}  considering an arbitrary tetragonal
distortion. In general, we write
\begin{align}
\hat{\mathcal{O}}^{L_\mu} =&U\ a_{\mu,U}\tau^{z}+Q_{2}\ a_{\mu,Q_{2}}\tau^{x}+Q_{3}\ a_{\mu,Q_{3}}\tau^{z}\nonumber \\
 & +\underset{a=x,y,z}{\sum}T_{a}\ a_{\mu,T_{a}}s^{a}\tau^{y}\nonumber \\
 & +\underset{a=x,y,z}{\sum}P_{a}\left(a_{\mu,P_{a}}s^{a}+b_{\mu,P_{a}}s^{a}\tau^{x}+c_{\mu,P_{a}}s^{a}\tau^{z}\right),\label{eq:RIXS scattering distortion}
\end{align}
where $\mu=2,3$ for the $L_{2,3}$ edge and we use the polarization factors given by Eq.
(\ref{eq:pol fac}). The above equation  corresponds to the projection
of the operators listed in Ref. \cite{Kim2017} to the $B$ and $C$ 
states discussed in Appendix \ref{sec:Orbital-Physics}. In Table
\ref{tab:Coefficients} we show the explicit values of the coefficients
in terms of the angle parameter $\varphi$ in Eq. (\ref{phiangle}). We
also highlight the coefficients in the cubic limit, which were expressed
in Eqs. (\ref{O2R}) and (\ref{O3R}). 

\bibliographystyle{apsrev4-1}
\bibliography{article2}

\begin{thebibliography}{73}
\bibitem [{com()}]{comentario}%
  \BibitemOpen
  \href@noop {} {}\bibinfo {note} {{The most important difference between the
  ansatz with $v=u$ and $v=-u$ is in the width of the bands in the
  $(\theta^1,\theta^3)$ sector. However, the difference between the two results
  for the ground state energy is rather small. The case $v=u$ studied in this
  work is the one that gives the lowest energy for Hamiltonian (1) at the
  mean-field level.}}\BibitemShut {Stop}%
\end{thebibliography}

\end{document}